%% file: vf-root.tex
\documentclass[journal]{IEEEtran}
\usepackage{amsmath,amsfonts}
\usepackage{algorithmic}
\usepackage{array}
\usepackage[caption=false,font=normalsize,labelfont=sf,textfont=sf]{subfig}
\usepackage{textcomp}
\usepackage{stfloats}
\usepackage{url}
\usepackage{verbatim}
\usepackage{graphicx}
\usepackage{cite}
\input{preamble_new.tex}
\usepackage[export]{adjustbox}

\renewcommand{\arraystretch}{1.5}
\usepackage{threeparttable}
\usepackage{multirow}

\usepackage{mdframed}
\definecolor{frame}{RGB}{200,200,255}

\mdfdefinestyle{theoremstyle}{%
	linecolor=frame,linewidth=1pt,%
	frametitlerule=true,%
	frametitlebackgroundcolor=gray!20, innertopmargin=\topskip,}

\mdtheorem[style=theoremstyle]{definition}{Hydraulic Dynamics}
\mdtheorem[style=theoremstyle]{definitionq}{Quality Dynamics}

\mdtheorem[style=theoremstyle]{definitionhc}{Decoupled Hydraulic Control}

\mdtheorem[style=theoremstyle]{definitionqhc1}{Quality-Aware Hydraulic Control --- \textit{Rank-Informed}}

\mdtheorem[style=theoremstyle]{definitionqhc2}{Quality-Aware Hydraulic Control --- \textit{Energy-Driven}}

\mdtheorem[style=theoremstyle]{definitionqhc3}{Quality-Aware Hydraulic Control --- \#3}

\usepackage{titlesec} 
\titlespacing*{\section}{0pt}{*0.7}{*0.3} 
\titlespacing*{\subsection}{0pt}{*0.7}{*0.3} 
\usepackage{tikz}
\usetikzlibrary{shapes.geometric, arrows.meta, positioning, calc}

\definecolor{Sub1purple}{RGB}{200,200,255}
\definecolor{Sub2yellow}{RGB}{251,234,206}
\definecolor{Sub3green}{RGB}{215,238,210}
\definecolor{BlueEdge}{RGB}{70,100,160}

\tikzstyle{startstop} = [
ellipse, draw=BlueEdge, thick,
minimum height=1.0cm, minimum width=1cm,
align=center, font=\normalsize, fill=white
]
\tikzstyle{parallelogram} = [
trapezium, trapezium left angle=70, trapezium right angle=110,
draw=BlueEdge, thick,
minimum width=1cm, minimum height=1cm,
align=center, font=\normalsize, fill=Sub1purple
]
\tikzstyle{process} = [
rectangle, draw=BlueEdge, thick,
minimum width=1cm, minimum height=1cm,
align=center, font=\normalsize, fill=Sub3green
]
\tikzstyle{decision} = [
diamond, aspect=2.3, draw=BlueEdge, thick,
minimum width=1cm, minimum height=1cm,
align=center, inner sep=0pt, font=\normalsize, fill=Sub2yellow
]
\tikzstyle{connector} = [
circle, draw=BlueEdge, thick,
minimum size=0.7cm, font=\normalsize
]
\tikzstyle{arrow} = [->, >=stealth, thick]

\begin{document}
\title{Quality-Aware Hydraulic Control in Drinking Water Networks via Controllability Proxies}
\author{Salma M. Elsheri$\text{f}^{\dagger, \P, \ast \ast}$, Mohamad H. Kazm$\text{a}^{\dagger}$, and Ahmad F. Tah$\text{a}^{\dagger}$
	\thanks{$^\dagger$Department of Civil and Environmental Engineering, Vanderbilt University, Nashville, TN, USA. Emails: salma.m.elsherif@vanderbilt.edu, mohamad.h.kazma@vanderbilt.edu, ahmad.taha@vanderbilt.edu}
	\thanks{$\P$Secondary appointment: Department of Irrigation and Hydraulics Engineering, Faculty of Engineering, Cairo University.}
	\thanks{$^{\ast \ast}$Corresponding author.}
	\thanks{This work is supported by the National Science Foundation under Grant 2151392.}
}
\maketitle

\begin{abstract}
	The operation of water distribution networks simply aims at efficiently delivering consumers adequate water while maintaining safe water quality (WQ). However, this process entails a multi-scale interplay between hydraulic and WQ dynamics evolving spatio-temporally within such a complex infrastructure network. While prior research has addressed the hydraulic optimization problem and WQ regulation as decoupled or coupled, they often overlook control-theoretic guided solutions. This paper takes a novel approach by investigating the coupling between hydraulic and WQ dynamics from a control networks perspective. We propose a quality-aware control framework that embeds WQ controllability metrics into the network-level pump scheduling problem, acknowledging the direct influence of system hydraulics on WQ controller behavior. We examine the trade-offs between pump control energy cost and WQ performance across various network sizes and scenarios. Our results showcase how network topology, hydraulic constraints, and WQ metrics jointly impact optimal pump schedules and, accordingly, the achievable level of WQ regulation, offering insights into designing efficient control strategies for water infrastructure networks governed by interdependent dynamics.
\end{abstract}


\section{Introduction}
\IEEEPARstart{T}{he} real-time management and operation of water distribution networks (WDNs) has been, and remains, one of the most researched topics in the field of water systems. The objective is to fulfill consumers' and end-users' needs and deliver clean water in a cost-effective manner while meeting water quality (WQ) mandates. Taking into account network-wide topology, operational flows, pressures, and WQ, achieving the aforementioned objective involves balancing multiple system requirements. These include minimizing the costs associated with meeting water demand and maintaining adequate pressure levels~\cite{Lai2021}, while satisfying quality standards under adequate disinfectant usage~\cite{xieNonlinearModelPredictive2015}.

For decision-makers (i.e., water utilities), these objectives pose challenges that must be addressed using control algorithms tailored for water distribution networks (WDNs). The regulation and control of WDNs primarily revolve around two aspects: quantity and quality.  
\textit{(i) Quantity}---the energy required to operate pumps constitutes the main component of the operational cost to be minimized while achieving the target head levels and flows \cite{puleoMultistageLinearProgramming2014}.  
\textit{(ii) Quality}---the objective of the control problem is to maintain desired disinfectant levels throughout the WDN with minimal injections at treatment plants and chlorination booster stations \cite{fisherFrameworkOptimizingChlorine2018}. We note that the operation of WDNs requires the consideration of a variety of quality parameters (e.g., turbidity, pH levels, and disinfectant residuals). However, among these parameters, disinfectant residuals stand out as a crucial indicator of the actual state of WQ. These residuals also present both modeling and computational challenges when optimizing their concentrations and injection strategies. That being said, our paper focuses on disinfectant modeling and control. In the remainder of this paper, WQ dynamics refers to disinfectant dynamics.

The vintage approach for studying WQ dynamics starts with first running hydraulic simulations that generate the pump schedules with their resulting heads and directional flows within a WDN. This is then followed by investigating WQ via booster station control. The reason for this decoupled approach is due to the difference in time scales between the hydraulic and WQ dynamics. To that end, this paper pursues a new approach: jointly investigating quality-quantity control via a unified approach based on network systems science and WQ controllability measures. Controllability, in this context, refers to the ability to effectively steer, regulate, and maintain disinfectant levels within the network to consistently meet the established water health standards.  In short, the paper attempts to answer the following research questions: \textit{Can quality-aware pump control significantly improve water quality dynamics in WDN? What are the quality controllability metrics that can be appended to a  hydraulic control problem? When is it meaningful to integrate the time-scales of quality and quantity?} To the best of our knowledge, this is the first attempt to explore this particular topic from a network systems and control-theoretic perspective.
\subsection{Relevant literature on WQ and hydraulic control}
The literature on the regulation and control of WDNs to deliver clean water to the end-users is rich and briefly summarized next. The literature on this topic is divided into \textit{(i)} studies focusing on determining the optimal operational settings for pumps and/or control valves to attain the desired water flows and levels specified by network topology and characteristics, and consumers demands; \textit{(ii)} studies that cover the regulation of WQ dynamics to ensure meeting the standard disinfectant (i.e., chlorine) residuals throughout the network while minimizing the source and booster stations' injections; and \textit{(iii)} studies that propose integrating/coupling controlling water quantity and quality, therefore considering the interdependencies between them. In this section, we survey these areas and afterward, we end this section by highlighting the research gaps that drive the contributions of this work. Tab.~\ref{tab:related_work_comparison} summarizes the surveyed literature on control strategies in WDNs. We note that the joint control approaches do not consider WQ controllability metrics, a gap in the literature.

\noindent \textit{\textbf{Hydraulic Control}.} \; The operational control of WDNs depends on several factors including network topology, demand cycle, tank dynamics, head loss model, pumps type, valves type, etc. In addition, the resulting control problem is inherently nonlinear and nonconvex in nature. Many studies have covered some or many of the aforementioned factors and handled the nonlinearity and nonconvexity under different frameworks. Studies \cite{puleoMultistageLinearProgramming2014,bonvinExtendedLinearFormulation2019} perform system linearization and apply linear programming (LP) to obtain optimal pump scheduling. Study  \cite{bonvinPumpSchedulingDrinking2021} determines the optimal scheduling by the means of relaxation and linear programming branch and bound (BnB). Similarly, the study in~\cite{singhOptimalSchedulingWater2019} relaxes the hydraulic constraints into second-order cone constraints with penalty terms, thereby enabling second-order cone programming (SOCP). On the other hand, studies~\cite{kangRealtimeOptimalControl2014a,wangNonlinearEconomicModel2017,wangMinimizingPumpingEnergy2021} solve nonlinear nonconvex problem using variations of methods including genetic algorithm (GA) and mixed-integer nonlinear programming (MINLP). Lastly, study \cite{wangRecedingHorizonControl2020} applies geometric programming-based model predictive control (MPC) algorithms which turn the problem into a convex continuous optimization problem. Many of these studies have compared their results with built-in tools in hydraulics solvers (e.g., EPANET's built-in rule-based control \cite{marchiPumpOperationOptimization2017})---refer to \cite{mala2017lost,reisReviewOperationalControl2023} for detailed review and analysis of the literature on this topic. 

\noindent \textit{\textbf{Quality Control}.} \;
A plethora of optimization-based approaches have been used to solve the WQ control problem including: LP with the objective of minimizing chlorine injections \cite{boccelliOptimalSchedulingBooster1998}, mixed-integer linear programming (MILP) with the allocation of booster stations as a decision variable \cite{trybyFacilityLocationModel2002}, and GA with a constraint on the formulation of disinfectant by-products \cite{oharOptimalDesignOperation2014}. Conversely to these studies, study \cite{wangHowEffectiveModel2021a} proposes applying MPC to an explicit representation of the single-species WQ model that guarantees network-wide control. More on that approach, study \cite{elsherifComprehensiveFrameworkControlling2024} utilizes different techniques to implement real-time MPC to nonlinear multi-species WQ dynamics---a framework that covers controlling chlorine levels in WDNs under abnormal conditions including contamination events. 
It is worth noting that these studies rely on the assumption that the system's hydraulics are pre-computed.

\begin{table*}[t]
	\fontsize{9}{9}\selectfont
	\centering 
	\caption{The type of control strategies (hydraulic, quality, and joint) covered by the research works in the literature and this paper.}
	\label{tab:related_work_comparison}
	\vspace{-0.2cm}
	\renewcommand{\arraystretch}{1.3}
	\resizebox{\linewidth}{!}{%
		\begin{tabular}{l|c|c|c|c|c|c|l}
			\midrule \hline
			\makecell{\textbf{Research} \\ \textbf{Work}} & 
			\makecell{\textbf{Hydraulic} \\ \textbf{Control}} &
			\makecell{\textbf{Quality} \\ \textbf{Control}} &
			\makecell{\textbf{Joint} \\ \textbf{Control}} & 
			\makecell{\textbf{Model} \textbf{Type}} & 
			\makecell{\textbf{Optimization} \\ \textbf{Method}} & 
			\makecell{\textbf{WQ Controllability} \\ \textbf{Metric}} &
			\makecell{\textbf{Control Framework}} \\ \hline
			
			\cite{puleoMultistageLinearProgramming2014, bonvinExtendedLinearFormulation2019, bonvinPumpSchedulingDrinking2021, singhOptimalSchedulingWater2019} 
			& \checkmark & -- & -- & Linear / Relaxed & LP / BnB / SOCP & -- & No quality control \\ \hline
			
			\cite{kangRealtimeOptimalControl2014a, wangNonlinearEconomicModel2017, wangMinimizingPumpingEnergy2021, wangRecedingHorizonControl2020} 
			& \checkmark & -- & -- & Nonlinear & GP / MINLP / MPC & -- & No quality control \\ \hline
			
			\cite{boccelliOptimalSchedulingBooster1998, trybyFacilityLocationModel2002, oharOptimalDesignOperation2014} 
			& -- & \checkmark & -- & Linear / Mixed-Integer & LP / MILP / GA & -- & Assumes pre-computed hydraulic conditions \\ \hline
			
			\cite{wangHowEffectiveModel2021a, elsherifComprehensiveFrameworkControlling2024} 
			& -- & \checkmark & -- & Nonlinear & MPC & -- & Assumes pre-computed hydraulic conditions\\ \hline
			
			\cite{sakaryaOptimalOperationWater2000a} 
			& -- & -- & \checkmark & Nonlinear & NLP & -- & Chlorine residual levels as constraint\\ \hline
			
			\cite{ostfeldConjunctiveOptimalScheduling2006} 
			& -- & -- & \checkmark & Nonlinear & GA & -- & Multi-objective optimization \\ \hline
			
			\cite{drewaMODELPREDICTIVECONTROL2007} 
			& -- & -- & \checkmark &  Nonlinear & MPC and GA & -- & Multi-objective optimization \\ \hline
			
			\cite{xieNonlinearModelPredictive2015} 
			& -- & -- & \checkmark &  Nonlinear & MPC & -- & Two-level optimization: pump control and chlorine injection \\ \hline
			
			\cite{abdallahFastPumpScheduling2019} 
			& -- & -- & \checkmark & Nonlinear & Goal Programming & -- & Multi-objective optimization\\ \hline
			
			\textbf{This Paper} 
			& -- & -- & \checkmark &  Nonlinear & NLP with MPC & \checkmark & Pump scheduling with explicit WQ controllability measures\\ 
			\toprule \bottomrule
	\end{tabular}}
\end{table*}
\setlength{\textfloatsep}{5pt}

\noindent \textit{\textbf{Joint Quality-Quantity Control}.} \;
Several studies have investigated integrating both the quantity and quality control problems by implicitly and/or explicitly incorporating one or more quality control aspects within the quantity control framework, or by turning them into one augmented formulation. In \cite{sakaryaOptimalOperationWater2000a}, the authors use nonlinear programming (NLP) to solve the pump operation problem that accounts for disinfectant's residuals in the constraints. On the other hand, study \cite{ostfeldConjunctiveOptimalScheduling2006} formulates a dual quality-quantity optimization problem with a single augmented objective that concatenates minimizing the energy cost and maximizing system's protection by maximizing the injected chlorine dose. This study utilizes a genetic algorithm to solve the optimal problem that is based on the two conflicting objectives. The posed problem is optimally solved while conveying the existence of trade-offs within the solution. Similarly, authors in \cite{drewaMODELPREDICTIVECONTROL2007} propose applying a genetic MPC algorithm to the coupled control problem and compare the results with real data records of a specific network, which shows cost reduction. The authors in \cite{xieNonlinearModelPredictive2015} utilize a nonlinear MPC integrated optimizer. The control procedure proposed is divided into two levels: an upper-level controller responsible for determining optimized pump schedules while satisfying constraints on chlorine residuals, and a lower-level controller that computes optimized chlorine injections. More recently, study \cite{abdallahFastPumpScheduling2019} solves a two-objectives pump scheduling problem by means of goal programming. The first objective focuses on implicitly achieving the required chlorine residuals by minimizing the active dynamics in storage components, while the other objective aims towards minimizing the energy cost. 

Many of the studies cited earlier on this topic solve a fully coupled quantity and quality control problem by concatenating the objectives of each problem into a single-objective or multi-objective formulation. This approach results in trade-offs between these objectives, highlighting the necessity for a thorough analysis of these trade-offs. On the other hand, incorporating a constraint on WQ, whether implicitly or explicitly, into the system's operational scheduling control problem does not necessarily guarantee the achievement of a certain level of controllability by booster stations or reachability of the desired final states. In other words, the notion of optimizing a pumping schedule while attaining a certain level of WQ controllability, based on a closed-form formulation of the system's hydraulics and quality, has not been attempted or investigated---a gap that is addressed in this paper.

\subsection{Paper Contributions}\label{sec:PaperCont}
The paper's objective is to investigate the feasibility and applicability of maintaining a certain level of control for the WQ model while computing the optimal hydraulic setting. This implies solving an augmented water network operational control problem that accounts for enhancing WQ from a control-theoretic perspective. Herein, we refer to the control problem which focuses solely on hydraulics as the \textit{decoupled problem}, while the joint quality-quantity problem as the \textit{coupled problem}. The paper contributions are as follows. 
\begin{itemize}[leftmargin=*]
	\item Water quality systems are inherently complex and mostly not fully controllable. That is, we investigate the effect of changing hydraulics on the WQ controllability. This is measured by employing different quantitative metrics which allow us to judge the WQ system controllability from different energy-related perspectives. Eventually, we judge the applicability and validity of these metrics on the case-oriented application.
	\item We formulate an augmented operational pump scheduling control problem---the coupled problem---in a way that conserves a certain level of WQ controllability. This level is predetermined depending on the investigation results performed as outlined in the previous point.   
	\item A wide comparison between the decoupled and coupled control problems through various numerical case studies is performed.
\end{itemize}

\noindent \textit{\textbf{Paper Organization.}} \;  We present the hydraulic and water quality models in Section~\ref{sec:HydWQModel}. In Section~\ref{sec:WQControlMetr}, we introduce controllability metrics and assess their applicability to our application. We then formulate the pumps scheduling optimization problem with and without the integration of WQ controllability preservation in Section~\ref{sec:HydOptProb}. 
We compare the decoupled and coupled problems through different case studies in Section \ref{sec:CaseStudies}. The paper's conclusions, limitations, and future work are discussed in Section \ref{sec:ConcRecom}.

\section{The Hydraulic and Quality Modeling of WDNs}\label{sec:HydWQModel}
We model WDNs as a directed graph, denoted as, $\mathcal{G} = (\mathcal{N},\mathcal{L})$.  The set $\mathcal{N}$ defines the nodes and is composed of subsets $\mathcal{N} = \mathcal{J} \cup \mathcal{T} \cup \mathcal{R}$ where the sets $\mathcal{J}$, $\mathcal{T}$, and $\mathcal{R}$ are the collections of junctions, tanks, and reservoirs. The set of links is defined as $\mathcal{L} \subseteq \mathcal{N} \times \mathcal{N}$, such that $\mathcal{L} = \mathcal{P} \cup \mathcal{M} \cup \mathcal{V}$ is composed of sets $\mathcal{P}$, $\mathcal{M}$, and $\mathcal{V}$ representing the collection of pipes, pumps, and valves. In each network, $n_\mathrm{L}$ and $n_\mathrm{N}$ represent the numbers of links and nodes. Specifically, network nodes include a number of $n_\mathrm{R}$ reservoirs, $n_\mathrm{J}$ junctions, and $n_\mathrm{TK}$ tanks. The total number of links, $n_{\mathrm{L}}$, can be written as the summation of $n_\mathrm{M}$, $n_\mathrm{V},$ and $n_\mathrm{P}$, representing the numbers of pumps, valves, and pipes, respectively. 

We consider two entities which comprise WDNs modeling (hydraulic and WQ models) with different numbers of states and representations. In the sequel, we formulate the state-space representation of both the hydraulic and WQ models for a given WDN, while succinctly summarizing the governing equations for each in Appendices~\ref{sec:hyd_model} and~\ref{sec:WQModel}. It is worth mentioning that the hydraulic time-step $\Delta t_{\mathrm{H}}$ is different than the WQ one $\Delta t_{\mathrm{WQ}}$. The hydraulic time-step is taken to be within an hourly scale to reflect the patterned demand, while the WQ time-step is chosen between minutes and seconds to allow for a stable and accurate numerical simulation \cite{seyoumIntegrationHydraulicWater2017}. Variable $t$ represents a specific time in the simulation period $[0,T_s]$ and it is updated incrementally by $\Delta t_{\mathrm{WQ}}$ within each $\Delta t_{\mathrm{H}}$ until reaching the end of the simulation period.

\subsection{Hydraulics in state-space form}\label{sec:Hyd-SSRep}
The detailed hydraulic modeling of the water network components (reservoirs, tanks, pipes, junctions, pumps, and valves) is presented in Appendix~\ref{sec:hyd_model}. The full hydraulics model explained therein can be written in the form of nonlinear difference algebraic equations (NLDAE) as expressed in \eqref{equ:HydSS}. In these equations, we collect the system state variables and inputs in vectors of appropriate dimensions as follows: heads at tanks in $\vw \in \mathbb{R}^{n_\mathrm{TK}}$; heads at junctions in $\vl \in \mathbb{R}^{n_\mathrm{J}}$; flows at pipes, pumps, and valves in $\vz \in \mathbb{R}^{n_\mathrm{L}}$; and relative speed of pumps $\vs\in \mathbb{R}^{n_\mathrm{P}}$. The vector $\m{\Omega} \in \mathbb{R}^{n_\mathrm{J}}$ encapsulates the junctions' demands which are considered predetermined in our study.

\begin{definition*}[\textbf{Hyd-NLDAE}]
	\vspace{-0.5cm}	\begin{subequations}\label{equ:HydSS}
		\begin{align}
			\vw (t+\Delta t_{\mathrm{H}}) &= {\mA_{\mathrm{H}}} \vw (t) + {\mB_{\mathrm{H}}} \vz (t),\label{equ:HydSSa} \\
			\boldsymbol{0}_{n_\mathrm{J}} &= \mE_z \vz(t) + \mE_\Omega {\m{\Omega}}(t),\label{equ:HydSSb} \\
			\boldsymbol{0}_{n_\mathrm{P}+n_\mathrm{M}+n_\mathrm{V}} &= \mE_w \vw(t) + \mE_l \vl(t) + {\m{\Psi}}(\vz,\vs),\label{equ:HydSSc}
		\end{align}
	\end{subequations}
\end{definition*} 
where $\{\mA, \mB, \mE\}_{\bullet}$ are constant matrices that depend on the network's topology, components' characteristics, and hydraulic parameters. Additionally, $\m{\Psi}(\cdot)$ gathers the nonlinear terms in \eqref{eq:PipeLosses}, \eqref{eq:PumpLosses}, and \eqref{eq:ValveLosses}, and $\boldsymbol{0}_{n}$ is a zero vector of size $n$.

\subsection{Water quality in state-space form}\label{sec:WQ-SSRep}
The WQ model (Appendix~\ref{sec:WQModel}) allows us to trace the disinfectant concentrations throughout the network components. The evolution of chlorine concentrations follows the conservation of chemical mass, transport, and single-species reaction and decay models. In each component, we represent the chlorine concentration as $c$ with a superscript for the component symbol.  Additionally, we highlight the hydraulics variables (e.g., velocities, flows, volumes, etc.) in \textcolor{violet}{violet} whenever they appear in the WQ model. The WQ model described in Appendices~\ref{sec:ConvMass}, \ref{sec:TranspReact}, and \ref{sec:SSmodel} can be formulated as the following linear difference equations (LDEs): 
\begin{definitionq*}[\textbf{WQ-LDE}]
	\vspace{-0.4cm}
	\begin{align}\label{equ:WQSS}
		\vx (t+\Delta t_{\mathrm{WQ}}) &= \textcolor{violet}{\mA_{\mathrm{WQ}}(t)} \vx (t) + \textcolor{violet}{\mB_{\mathrm{WQ}}(t)} \vu (t), \\
		\vy (t) &= \mC_{\mathrm{WQ}}(t) \vx (t),
	\end{align}
\end{definitionq*}
where vector $\vx(t) := \{\vc^\mathrm{R}(t), \; \vc^\mathrm{J}(t), \; \vc^\mathrm{TK}(t), \; \vc^\mathrm{P}(t), \; \vc^\mathrm{M}(t), \;$ $\vc^\mathrm{V}(t)\} \in \mathbb{R}^{n_x}$ depicts the concentrations of chlorine in the entire network and the total number of states $n_x=n_\mathrm{R}+n_\mathrm{J}+n_\mathrm{TK}+\sum_{i=1}^{n_\mathrm{P}}n_{s_i}+n_\mathrm{M}+n_\mathrm{V}$; vector $\vu(t) \in \mathbb{R}^{n_{u}}$ represents the dosages of injected chlorine; vector $\vy(t) \in \mathbb{R}^{n_{y}}$ denotes the sensor measurements of chlorine concentrations at specific locations in the network. The state-space matrices $\{\mA_{\mathrm{WQ}}, \mB_{\mathrm{WQ}}, \mC_{\mathrm{WQ}}\}$ are all time-varying matrices that depend on the network topology and parameters, hydraulic parameters, decay rate coefficients for the disinfectant, and booster station and sensor locations. It is customary to assume that these matrices evolve at a slower pace than the states $\vx(t)$ and control inputs $\vu(t)$. This is due to the slower evolution of hydraulic variables, such as flows and heads used in constructing the WQ system matrices, compared to the states and inputs related to chlorine concentrations.

\section{Water Quality Controllability Proxies}\label{sec:WQControlMetr}
In this section we introduce the notions of controllability for linear WQ dynamics. We define controllability measures that allow us to qualitatively and quantitatively evaluate the controllability of the developed WQ system~\eqref{equ:WQSS}. These metrics are assayed based on their applicability to WQ models and their suitability to the optimal pump scheduling problem.

We consider the notion of WQ system controllability and relate it to the hydraulic pump scheduling problem formulated in Section~\ref{sec:HydOptProb}. From a control-theoretic perspective, controllability is the ability to steer a system from initial states $\m{x}_{o}:=\m{x}_{0}$ to $\m{x}_{s}:=\m{x}_{T_s}$ by some input $\m{u}(t)$~\cite{kalmanMathematicalDescriptionLinear1963}. That is, the goal is to be able to steer complex dynamical systems to a desired state or trajectory. Specifically, for WQ control we want to maintain chlorine concentrations within certain levels.

\begin{mydef}\label{def:control}
	A linear system [e.g., the WQ system expressed as \eqref{equ:WQSS}] is controllable if for any finite time interval $[0,T_s]$ and for any initial state $\vx_{o} \in \mathbb{R}^{n_{x}}$, the initial state $\vx_{o}$ can be steered or driven to a final state $\vx_{T_s} \in \mathbb{R}^{n_{x}}$ for some input $\vu(t)$ under the specified time interval.
\end{mydef}

That being said, the dynamic linear system \eqref{equ:WQSS} is said to be controllable if only if the controllability matrix for $N_s= \frac{T_s}{\Delta t_{\mathrm{WQ}}}$ time-steps given as
\begin{multline}\label{eq:control_matrix}
	\mathcal{C}_{N_{s}} := \{\begin{matrix}
		\mB_{\mathrm{WQ}}, \;\; \mA_{\mathrm{WQ}} \mB_{\mathrm{WQ}}, \;\; \mA_{\mathrm{WQ}}^{2} \mB_{\mathrm{WQ}}, \end{matrix} \\   \begin{matrix} \dots, \;\; \mA_{\mathrm{WQ}}^{N_s-1} \mB_{\mathrm{WQ}}
	\end{matrix}\} \in \mathbb{R}^{n_{x}\times N_{s}n_{u}},
\end{multline}
is full row rank, i.e, $\mr{rank}(\mathcal{C}_{N_{s}}) = n_x$~\cite{hespanhaLinearSystemsTheory2018a}. Without loss of generality as we assume that  $N_{s}n_{u} > n_{x}$. This is known as Kalman's rank condition~\cite{kalmanMathematicalDescriptionLinear1963}. However, matrix rank is a generic property that might lead to similar values depending on the relations between the variables; it therefore is informative in a qualitative sense but fails to indicate how controllable the system is under many cases and various scenarios. 

For the WQ dynamics~\eqref{equ:WQSS}, full row rank of $\mathcal{C}_{N_{s}}$ seldom occurs---this is due to the complexity and the high dimensionality of the system. 
With that in mind, it is more practical to consider quantitative measures of controllability which, unlike the aforementioned rank metric, are able to reflect the difficulty in controlling the WQ system. 

To that end, the notion of control energy $\mathcal{E}(\m{A}_\mathrm{WQ},\m{B}_\mathrm{WQ},N_s,\m{x}_{o}\rightarrow\m{x}_{T_{s}})$ is introduced to quantify the energy needed to steer the system from $\m{x}_{o}$ to $\m{x}_{T_{s}}$. Ideally, we want to minimize the energy required to control the system. The concept of energy-related control depends on the application. For the case of WQ control, it is related to the amount of chlorine needed to be injected into the system to keep a desired chlorine level at the network's components. 

Metrics related to the input energy are based on the controllability Gramian $\m{W}_{c}(\m{A}_\mathrm{WQ},\m{B}_\mathrm{WQ},N_s):= \m{W}_{c} \in \mathbb{R}^{n_{x}}$ that is defined for $N_{s}$ as the sum of matrices pair $\m{A}_\mathrm{WQ}$ and $\m{B}_\mathrm{WQ}$ as
\begin{equation}\label{equ:control_gram}
	\hspace{-0.5cm}	\m{W}_{c} := \sum_{\tau=0}^{N_s-1}\m{A}_\mathrm{WQ}^{\tau}\m{B}_\mathrm{WQ} \m{B}_\mathrm{WQ}^{\top}(\m{A}_\mathrm{WQ}^{\top})^{\tau} = 
	\mathcal{C}_{N_{s}}\mathcal{C}_{N_{s}}^{\top},
\end{equation}
where the controllability Gramian $\m{W}_{c}$, that is a positive semidefinite metrics, provides an energy-related quantification of controllability such that, $\mathcal{E} \propto \mr{trace}(\m{W}_{c})^{-1} $. We note here that $\m{W}_{c}$ is non-singular if the system is controllable after time ${T_{s}}$; otherwise, it is uncontrollable. 

\begin{myrem}\label{rem:rem1Metrices}
	The WQ system matrices $\m{A}_\mathrm{WQ}$ and $\m{B}_\mathrm{WQ}$ are {“time-varying”} throughout the simulation window due to changes in hydraulic dynamics. However, within each hydraulic time-step, they are considered {“time-invariant”}, as the hydraulic variables are not updated until the end of each hydraulic time-step.
\end{myrem}

In the literature~\cite{pasqualettiControllabilityMetricsLimitations2014a, summersSubmodularityControllabilityComplex2016a} a myriad of measures exist; these measures provide a scalar energy-related quantification of the controllability Gramian $\m{W}_{c}$. These measures include: $\mr{log}\mr{det}(\m{W}_{c})$, $\mr{trace}(\m{W}_{c})$, 
$\mr{rank}(\m{W}_{c})$, and minimum eigenvalue $\lambda_{\min}(\m{W}_{c})$. A discussion on the aforementioned measures is given as follows. 
\begin{itemize}[leftmargin=*]
	\item The $\mr{log}\mr{det}(\m{W}_{c})$ metric is proportional to the volumetric measure of the ellipsoid enclosing the set of states that can be reached with at most a unit control energy input. 
	\item The $\mr{trace}(\m{W}_{c})$ metric is inversely related to the average controllability energy in all directions of the state-space. 
	\item The $\mr{rank}(\m{W}_{c})$ metric quantifies the size of the controllable subspace.
	\item The $\lambda_{\min}(\m{W}_{c})$ metric is inversely related to the 
	control energy in the most difficult control direction. The smallest eigenvalue quantifies the worst-case direction that requires the largest amount of control energy.
\end{itemize}

In terms of WQ controllability, the above metrics have several interpretations. For instance, the $\mr{rank}$ metric can be interpreted as quantifying the extent to which an operator (i.e., booster station) can influence network components (extent of WQ control coverage). As such, the larger the $\mr{rank}$ of the controllability Gramian, the greater the number of network components where the chlorine injections have an effective influence on the residual concentrations within the specified time interval. The $\mr{trace}$ and $\mr{logdet}$ quantify the energy in all directions of the state-space. Thus, maximizing the control energy within a system signifies a greater capacity for the chlorine injections to impact the various states within the water network over the specified time interval. The $\lambda_{\mr{min}}$ indicates the largest energy needed, which is translated as chlorine injections, for a specific direction to influence its system states and steer them to the desired states. 

It is worth mentioning that the notion of controllable subspace is equivalent to the notion of \textit{reachable} subspace; this is related to the representation of the Gramian and its associated metrics \cite{fuhrmannWeakStrongReachability1972}. A system is said to be reachable in a specific state space if the subspace of all the reachable states from an initial state $\vx_{o}$ is equal to the whole state space. Attaining this property is important when controlling complex WQ dynamics. Our goal herein is to preserve and maintain a certain level of energy within a \textit{controllable and reachable subspace}. Thus, defining the controllable subspace is instrumental in measuring metrics such as $\mr{trace}$, $\mr{logdet}$, and $\lambda_{\min}$ when the system is not full rank; see Appendix~\ref{sec:reach_control} for the definition of controllable subspace.

\section{Hydraulic and Water Quality Driven Control}\label{sec:HydOptProb}
In this section, we formulate the real-time control and pump scheduling optimization problem to achieve the stated goal of preserving and maintaining WQ through incorporating the aforementioned energy driven metrics. To that end, we present the decoupled and coupled pump scheduling problems. 
\subsection{Decoupled pump scheduling problem}\label{sec:OptPumpAlone}
The objective of the decoupled pump scheduling program is to minimize the pump power consumption while being subjected to the system dynamics and functional constraints. We summarize the decoupled pump scheduling optimization derivation and formulation in Appendix~\ref{sec:de_pump_schedule}. Note that, modifications made to the pipe and pump dynamics are to ensure a convex formulation which in turn lead to the elimination of the nonlinear formulation in \eqref{equ:HydSS} introduced by \eqref{equ:HydSSc}. Henceforward, we incorporate these modifications in addition to the linear expressions in \eqref{equ:HydSS}, referring to this combination as Hyd-LDAE. The objective that is defined as minimizing cost of power consumption by pumps is defined by $\Pi(t)$ (Appendix~\ref{sec:de_pump_schedule}, Eq.~\ref{equ:PumpPowerActual}) and approximated as $\Pi_{\mathrm{App}}(t)$ (Appendix~\ref{sec:de_pump_schedule}, Eq.~\ref{equ:PumpPowerApp}). The decision variables are collected in vector $\m{\Upsilon}(t):=\{\vw(t), \vl(t), \vz(t), \m{\zeta}(t), \m{\omega}(t), \vs(t)\}$. Therefore, the final decoupled optimization problem is represented as a mixed-integer quadratic constrained quadratic problem (MIQCQP) for each hydraulic time-step:

\begin{definitionhc*}
	\vspace{-0.4cm}
	\begin{subequations}\label{eq:DecoupledHydOpt}
		\begin{align}
			\underset{\m{\Upsilon}(t)}{\mbox{minimize}} \hspace{1cm} & \Pi_{\mathrm{App}}(\m{\Upsilon}(t)) \Delta t_{\mathrm{H}} \\
			\begin{split}\label{eq:DecHydOptConst}
				\mbox{subject to} \hspace{1cm} & \mbox{\textbf{Hyd-LDAE}} \; \eqref{equ:HydSSa}, \eqref{equ:HydSSb}, \\
				& \eqref{equ:PipeConstraints}, \; \eqref{equ:AppPumpCurve}, \;   \eqref{equ:HydPhysConstraints}.
			\end{split}	
		\end{align}		
	\end{subequations}
\end{definitionhc*}

The constraints in~\eqref{equ:PipeConstraints} represent the pipe constraints, which include head loss within pipe segments and the flow equality constraint. Constraint~\eqref{equ:AppPumpCurve} is the pump curve constraint. Physical constraints on the head levels and flow values among the various network components are represented by~\eqref{equ:HydPhysConstraints}.

\subsection{Joint quality-aware pump control problem}\label{sec:OptPumpWQ}
The main results of this manuscript are presented in this section. Herein, we incorporate the discussed WQ controllability Gramian (WQ-CG) metrics in the previously developed pump scheduling problem to formulate the coupled WQ controllability-guided pump scheduling optimization problem. The dependency of these metric calculations on the hydraulics settings is direct, yet complex and leads to highly nonlinear expressions---Appendix \ref{sec:App1} showcases the raised concern for the case of the Three-node network. In addition, multiple factors should be taken into consideration that reflect the general and specific characteristics of the operation of WDNs. These factors impose logical and physical constraints on the choice and purpose of controllability metric integrated into our problem. For instance, during the initial simulation time with zero initial chlorine concentrations throughout the network, we aim for a higher controllability Gramian rank along with high energy. This approach also applies to branched networks with numerous dead-ends to ensure chlorine concentrations remain within standard limits. Another scenario arises when we need to store water with sufficient chlorine concentrations in tanks during off-peak demand periods, ready for distribution when the tanks are in demand, supplying either specific network sections or the entire network. 

Formulating and solving a problem that takes these factors int account and achieves the desired level of WQ controllability presents several challenges. These challenges primarily arise from the high nonlinearity and complexity involved in defining these metrics. Note that, as emphasized in Remark \ref{rem:rem1Metrices}, the WQ system matrices are time-invariant within the same hydraulic time-step. Consequently, the controllability Gramian and associated metrics, as discussed, are also invariant over this period. However, the formulation of the WQ state-space matrices, and consequently, its controllability Gramian, depends on factors such as flow directions in each pipe and the number of segments into which it is discretized. Yet, these factors are to be determined by solving the problem itself. In response to these challenges, we propose following the approach developed herein to address these issues effectively.
\begin{figure}[t]
	\centering
	\scriptsize
	\resizebox{1\columnwidth}{!}{
		\begin{tikzpicture}[node distance=0.5cm and 0.8cm]
			\node[startstop] (start) {Start};
			\node[parallelogram, below=of start] (topology) {Network Topology,\\ Characteristics,\\ Physical Constraints};
			\node[parallelogram, below=of topology] (segments) {Specify Number of Segments for each\\ Pipe $i \in \mathcal{P}$};
			\node[process, below=of segments, fill=Sub1purple] (wc) {Build Controllability Gramian $\m{W}_{c} $};
			\node[parallelogram, below=of wc] (targets) {Target Nodes $\mathcal{T}$};
			\node[process, below=of targets, fill=Sub1purple] (wt) {Build and Approximate Target\\ Controllability Gramian $\m{\widetilde{W}}_{\mathcal{T}}$};
			\node[process, below=of wt, fill=white] (t0) {$t = 0$};
			\node[connector, below=of t0] (A1) {A};
			
			\node[connector, right=5.5cm of start] (A2) {A};
			\node[decision, below=of A2] (whileloop) {While $t \in [0, T_s]$};
			\node[decision, below=of whileloop] (proxy) {Controllability \\ Proxy};
			\node[process, below=1.0cm of proxy] (ranked) {
				Solve Quality-Aware Hydraulic\\ Control Problem --\\ \textit{\textbf{Rank-Informed}}};
			
			\node[process, right=1.3cm of proxy] (energy) {
				Solve Quality-Aware Hydraulic\\ Control Problem --\\ \textit{\textbf{Energy-Driven}}};
			\node[decision, below=of ranked] (feasible) {Feasible Solution};
			\node[process, right=1.8cm of feasible] (solve) {Solve Decoupled Hydraulic\\ Control Problem};
			\node[process, below=of feasible, fill=white] (advance) {$t = t + t_H$};
			
			\node[parallelogram, right=1.5cm of whileloop, text width=2.1cm] (schedule) {Optimal Pump\\ Schedule};
			\node[startstop, right=of schedule] (end) {End};
			
			\draw[arrow] (start) -- (topology);
			\draw[arrow] (topology) -- (segments);
			\draw[arrow] (segments) -- (wc);
			\draw[arrow] (wc) -- (targets);
			\draw[arrow] (targets) -- (wt);
			\draw[arrow] (wt) -- (t0);
			\draw[arrow] (t0) -- (A1);
			
			\draw[arrow] (A2) -- (whileloop);
			\draw[arrow] (whileloop) -- node[left, font=\small] {True} (proxy);
			\draw[arrow] (whileloop.east) -- node[above, font=\small] {False} (schedule);
			\draw[arrow] (schedule) -- (end);
			\draw[arrow] (proxy.south) -- node[left, font=\small] {Rank} (ranked.north);
			\draw[arrow] (proxy.east) -- node[above, font=\small] {Energy} (energy.west);
			\coordinate (mergepoint) at ($(ranked.south)!0.5!(feasible.north)$);
			\draw[arrow] (ranked) -- (feasible);
			
			\draw[arrow] (feasible.south) -- node[left, font=\small] {Yes} (advance.north);
			\draw[arrow] (energy.south) |- (mergepoint);
			\draw[arrow] (solve.south) |- (advance.east);
			\draw[arrow] (feasible.east) -- node[above, font=\small] {No} (solve.west);
			\draw[arrow] (advance.west) -- ++(-1.8,0) |- (whileloop.west);
		\end{tikzpicture}
	}
	\vspace{-0.5cm}
	\caption{Flowchart of the proposed quality-aware hydraulic control framework for WDNs.}\label{fig:FlowChart}
\end{figure}

First, we overcome the flow direction issue while formulating the $\mA_\mathrm{WQ}$ matrix by building it for both cases and use the introduced binary variables $\omega(t)$ for each pipe. These variables define which pipe piecewise-linearization segment is chosen and accordingly the flow direction. For each element of the matrix depending on the flow direction for a specific pipe, it is multiplied by the summation of half of the $\omega(t)$ variables representing the corresponding direction.  

In addition, as explained in Section \ref{sec:HydWQModel}, the number of segments defined for each pipe depends on the water velocities (i.e., hydraulics in the system and decision variables of our problem). This number of segments defines the WQ model dimensions for the simulation window and accordingly the dimensions of the $\mA_\mathrm{WQ}$ and $\mB_\mathrm{WQ}$ matrices. Yet, the hydraulics variables are to be determined by the problem through which we aim to account for the WQ controllability. To that end, we define the WQ system dimensions offline as a prior-control step that preserves WQ model stability and hydraulics applicable scenarios. We randomly generate pump speeds between 0 and 1 and solve the systems' flow and heads according to a variety of demands patterns. Then, we calculate the number of segments needed for each pipe to ensure a fulfilled stability condition. Finally, we define this number to be the minimum for each pipe out of all the scenarios. This approach guarantees that after solving the pump scheduling problem and obtaining actual operational hydraulic setting, the WQ model is stable and has been well-represented.

Given that our optimization problem is tailored for the specific purpose of enhancing dynamics within WDNs, we simplify the utilization of WQ controllability metrics within this context. This simplification is based on the characteristics of the dynamics inherent to these systems. These distinctive dynamics include: \textit{(i)} booster stations that are located only at nodes, \textit{(ii)} pipes that are discretized into a number of segments within the WQ dynamics simulation, thus ensuring controllability over such segments, and \textit{(iii)} in many scenarios, the objective is to obtain higher controllability coverage and/or energy to reach specific junctions/dead-ends that serve large areas and/or tanks scheduled for on-demand operation during various simulation intervals. To that end, we adopt the concept of \textit{target controllability} \cite{gokhaleOptimizingControllabilityMetrics2021a} given in Definition \ref{def:TargetControl}. Target controllability allows us to choose the desired target nodes and, accordingly, eliminates the large dimentionality issue associated with the WQ representation. In this case, the metrics are applied to the \textit{targeted controllability Gramian} $\boldsymbol{W}_{\mathcal{T}}=C_\mathcal{T} \m{W}_c C_\mathcal{T}^\top$. 

\begin{mydef}\label{def:TargetControl}
	A discrete linear system is said to be
	target controllable with respect to the target set $\mathcal{T} \subseteq \mathcal{G}$; $|\mathcal{T} |=n_{y_f}$ over time $[0,T_s]$, if for any final output $y_f(t) = C_\mathcal{T} x(t),$ $y_f \in \mathbb{R}^{n_{y_f}}$ and $\vx_{o} \in \mathbb{R}^{n_{x}}$, the initial state $\vx_{o}$ can be steered or driven to final state of the target nodes as  $y_f$ for some input $\vu(t)$ under the specified time interval. The output matrix $C_\mathcal{T}$ identifies the set of
	target nodes $\mathcal{T}$.
\end{mydef}

By integrating constraints on the target controllability Gramian to Eq. \eqref{eq:DecoupledHydOpt}, the coupled optimization problem is formulated in \eqref{eq:CoupledHydWQOpt} as a nonconvex nonlinear problem.
\begin{subequations}\label{eq:CoupledHydWQOpt}
	\begin{align}
		\hspace{-0.5cm}	\underset{\m{\Upsilon}(t)}{\mbox{minimize}}\;\;\; & \Pi_{\mathrm{App}}(\m{\Upsilon}(t)) \Delta t_{\mathrm{H}} - \Theta_1 {\rm{trace}}(\m{W}_{\mathcal{T}}(\m{\Upsilon}(t))) \notag \\ 
		& - \Theta_2 {\lambda_{\min}}(\m{W}_{\mathcal{T}}(\m{\Upsilon}(t))), \label{eq:DecHydOptConst_1}  \\
		\hspace{-0.5cm}	\mbox{subject to} \; \;\; & \mbox{\textbf{Hyd-LDAE}} \; \eqref{equ:HydSSa}, \eqref{equ:HydSSb}, \eqref{equ:PipeConstraints}, \eqref{equ:AppPumpCurve},   \eqref{equ:HydPhysConstraints}, \notag \\ & 
		{\rm{rank}}(\m{W}_{\mathcal{T}}(\m{\Upsilon}(t))) \geq n_{y_f}-1,  
	\end{align}		
\end{subequations}
where $\Theta_1$ and $\Theta_2$ are scaling factors. 

We choose the target nodes to be the ones listed in point \textit{(iii)} along with the first and last segment of the connecting pipes to assure the validation of the flow directions. However, although employing the targeted controllability Gramian reduces the number of constraints, the integration the WQ metrics results in nonlinear problem formulation. Nonlinear solvers can address these issues, but they demand significantly more runtime, which escalates exponentially as network size increases. Additionally, the presence of the "$\rm{rank}$" constraint imposes limitations on this formulation, as many solvers do not support constraints related to rank. Thus, we employ the following simplifications to accelerate the computational process while achieving the desired output. Some of these simplification strategies are suitable for relatively small systems, while others are tailored for larger networks. In addition, some preparation steps can be taken for all network sizes. 

First, we approximate the Gramian by eliminating the denominator in all values of $\mA_\mathrm{WQ}$ and $\mB_\mathrm{WQ}$ to only result in polynomial expressions in the Gramian and change its notation to $\m{\widetilde{W}}_{\mathcal{T}}$. This helps to distinguish the contribution of columns/rows in $\rm{rank}$ determination and in the $\rm{trace}$ and $\lambda_{\min}$ computations. Second, depending on the scenario under consideration, optimization problem \eqref{eq:CoupledHydWQOpt} can be modified to be rank-oriented to achieve desired controllability coverage or/and energy-oriented to achieve desired controllability energy. To solve the $\rm{rank}$ constraint issue, we employ the approach detailed in \cite{maiConvexMethodsRankConstrained2015}, which involves applying a convex relaxation of the rank constraint by using a nuclear norm penalty and specifying the required rank---formulated as Eq. \eqref{eq:CoupledHydWQOptRank}. On the other hand, the energy-oriented problem (Eq. \eqref{eq:CoupledHydWQOptEnergy}) is formulated to avoid the transformation needed to maximize the $\lambda_{\min}$ by maximizing the $\rm{trace}$ of the target controllability Gramian built for a specific direction.

\begin{definitionqhc1*}
	\vspace{-0.35cm}
	\begin{subequations}\label{eq:CoupledHydWQOptRank}
		\begin{align}
			\hspace{-0.4cm}	\underset{\m{\Upsilon}(t)}{\mbox{minimize}}\;\;\; & \Pi_{\mathrm{App}}(\m{\Upsilon}(t)) \Delta t_{\mathrm{H}} - \Theta_3 ||\m{\widetilde{W}}_{\mathcal{T}}(\m{\Upsilon}(t))||_{*},  \\
			\hspace{-0.4cm}	\mbox{subject to} \; \;\; & \mbox{\textbf{Hyd-LDAE}} \; \eqref{equ:HydSSa}, \eqref{equ:HydSSb}, \eqref{equ:PipeConstraints}, \eqref{equ:AppPumpCurve},   \eqref{equ:HydPhysConstraints}, \notag \\ & 
			l_r||(\m{\widetilde{W}}_{\mathcal{T}}(\m{\Upsilon}(t)))||_2  - {\rm{trace}}(\m{\widetilde{W}}_{\mathcal{T}}(\m{\Upsilon}(t)))  \leq 0,  
		\end{align}		
	\end{subequations}
\end{definitionqhc1*}
where $\Theta_3$ is scaling factor, $||\m{\widetilde{W}}_{\mathcal{T}}||_{*}$ is the nuclear norm, $||\m{\widetilde{W}}_{\mathcal{T}}||_{2}$ is the second norm, and $l_r$ is the desired rank of the target controllability Gramian.

\begin{definitionqhc2*}
	\vspace{-0.5cm}
	\begin{subequations}\label{eq:CoupledHydWQOptEnergy}
		\begin{align}
			\hspace{-0.45cm}	\underset{\m{\Upsilon}(t)}{\mbox{minimize}}\;\;\; & \Pi_{\mathrm{App}}(\m{\Upsilon}(t)) \Delta t_{\mathrm{H}} - \sum_{i=1}^{n}\Theta_i {\rm{trace}}(\m{\widetilde{W}}_{\mathcal{T}_i}(\m{\Upsilon}(t)))  \\ 
			\hspace{-0.55cm}	\mbox{subject to} \; \;\; & \mbox{\textbf{Hyd-LDAE}} \; \eqref{equ:HydSSa}, \eqref{equ:HydSSb}, \eqref{equ:PipeConstraints}, \eqref{equ:AppPumpCurve},   \eqref{equ:HydPhysConstraints}, \notag \\
			& {\rm{trace}}(\m{\widetilde{W}}_{\mathcal{T}_i}(\m{\Upsilon}(t))) \geq 0 \;\;\; \forall i=1, \dots, n.
		\end{align}		
	\end{subequations}
\end{definitionqhc2*}

The proposed framework is suitable for networks of all sizes. However, in scenarios where multiple controllability constraints need to be satisfied for large networks, the computational time can become demanding. That is, for these networks the controllability Gramian can be built for a defined \textit{important} path of the network. In such cases, we propose focusing on building the controllability Gramian for specific \textit{important} network paths. Depending on network characteristics, this path can be determined based on factors like high demand, water transfer between reservoirs and elevated tanks, or mainline locations before branching occurs. To that end, our approach is applicable for the entire network or parts of it.

It is worth mentioning that the final formulation is nonlinear and nonconvex, yet tackled by several solvers and the level of complexity and accordingly the runtime are determined by the scenario under consideration. Additionally, In some cases constraining the optimization problem to reach specific level of controllability results in infeasible problem. To address this, we put a condition on our formulation to reduce that level for such case and re-solve the problem successively until a minimum level of controllability is proven to be unattainable. Under such condition, we transition to solving the decoupled problem and focus on further improvements in subsequent time-steps. Fig. \ref{fig:FlowChart} summarizes the flowchart of the proposed framework.

After building the proposed framework, we discuss herein the expected results according to the pump status post-solving the decoupled problem \eqref{eq:DecoupledHydOpt} or any of the coupled problems \eqref{eq:CoupledHydWQOptRank} or \eqref{eq:CoupledHydWQOptEnergy}. Firstly, we emphasize that the flow is one-directional in pumps, which means that the pump only provides head increase. That is, while the pump is switched on,  the variables associated with the pump outputted from either control problems are the flow through the pump, its operating relative speed, and the difference in head between the downstream and upstream nodes. Nonetheless, while the pump is switched off, there are two possible scenarios of outputs that can be obtained:
\begin{enumerate}[leftmargin=*]
	\item The pump to have a positive flow with a zero head increase. This is considered a valid scenario as it can present a bypass link to the pump to allow water to flow from the upstream  to downstream nodes. As the pump is assumed to be a link with a really small and almost zero length, the bypass link can hold the same assumption and accordingly, the change in head to be zero is valid.
	\item The pump to have a zero flow, yet the head increase is positive. Practically, this head increase is called the shut-off head of the pump. However, achieving this condition practically is not feasible as there is no water flowing through the pump. Similarly to the first scenario, it is a valid scenario but reflects on a different operation setting. Typically, pumps are equipped with valves to prevent backflow, ensuring that all the head above the pump is dissipated in that valve. In this scenario, water flows towards the pump from the downstream node, and the head difference is lost in the valve. To validate this scenario, it is assumed that the node downstream of the pump is at an equal or higher elevation than the upstream node, eliminating the need for the head through the pump to incur a head loss under any circumstance.
\end{enumerate}
In both scenarios, the results of the decoupled or coupled problems may display a \textit{pseudo} pump relative speed, identified by examining the flow and head increase, and subsequently disregarded. After the pump schedules are computed for the simulation period, the WQ control problem can be solved by the appropriate chosen technology. In our paper, we apply the model predictive control (MPC) algorithm adopted in \cite{wangHowEffectiveModel2021a}. For brevity, we do not include the details and the derivation of the control problem and we refer the reader to the cited study reaching the final formulation. Note that, the WQ control problem is constrained by upper and lower bounds on the chlorine concentrations at all network components. These bounds are specified by EPA regulations to be between $\vx_{\min}=0.2$ mg/L and $\vx_{\max}=4$ mg/L~\cite{acrylamideNationalPrimaryDrinking2009}.  

\begin{figure}[t]
	\centering
	\includegraphics[width=0.49\textwidth]{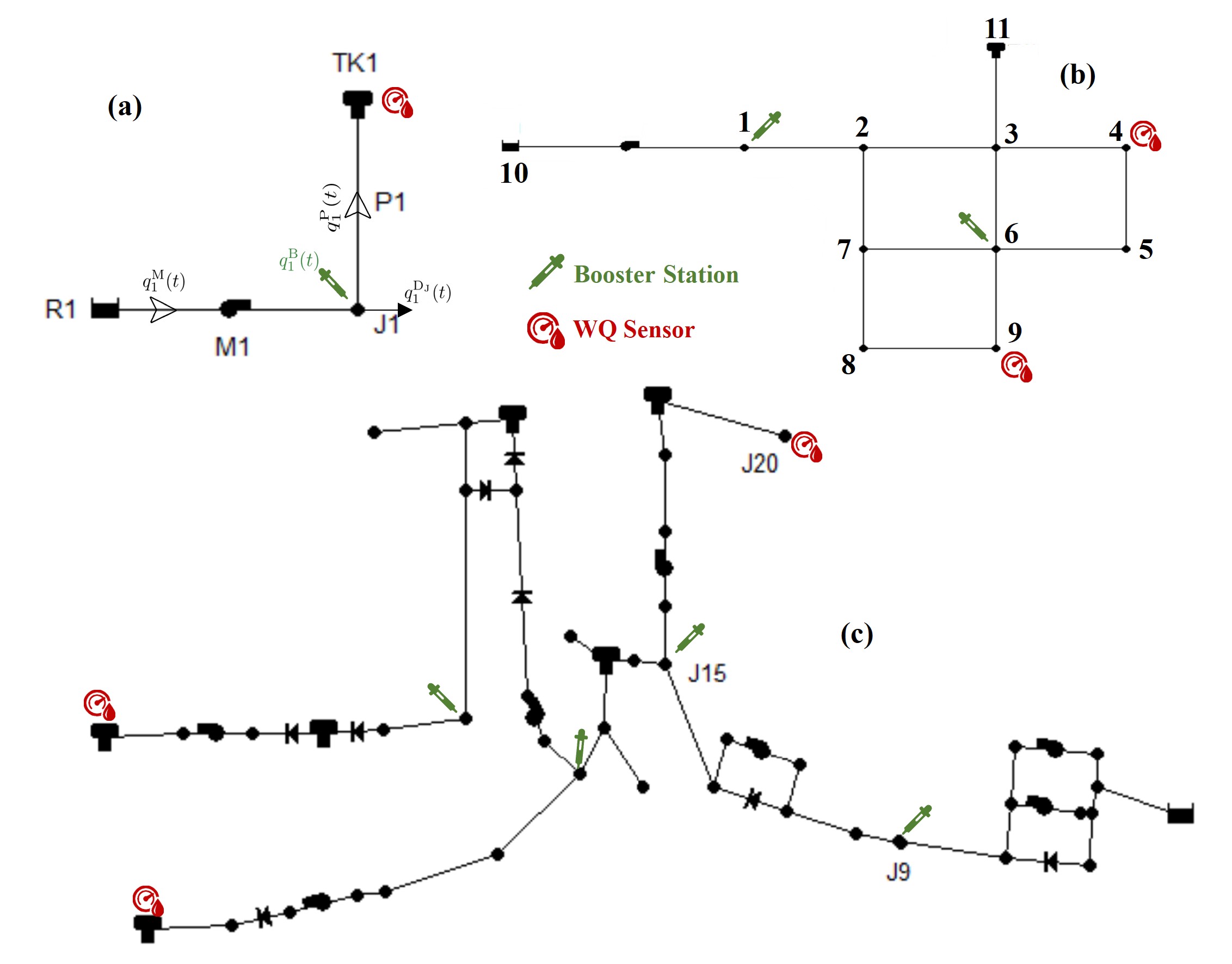}
	\vspace{-0.5cm}
	\caption{Networks under study and their topological layouts: (a) Three-node network, (b) Net1, and (c) Richmond skeleton network.}\label{fig:CaseStudy}
\end{figure}

\begin{figure}[b]
	\centering
	\includegraphics[width=0.49\textwidth]{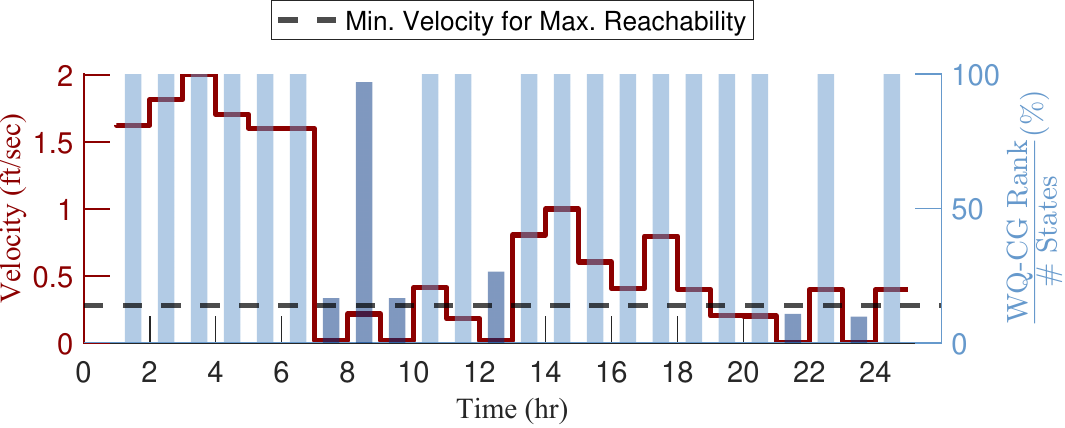}
	\vspace{-0.7cm}
	\caption{Flow velocity pattern in P1 of the Three-node network in comparison to the minimum required velocity to achieve maximum reachability of the flow along the pipe's length within the hydraulic time-step vs. the corresponding percentage (\%) of the system's WQ-CG rank out of the \# states in the subspace. Change in bars colors to highlight (darker shade) the windows where the WQ-CG is not full rank.}\label{fig:HydvsWQContrl}
\end{figure}

\section{Case Studies}\label{sec:CaseStudies}
In our study, we investigate the validity of our proposed control algorithms on three networks; Three-node network, Net1, and the Richmond skeleton network \cite{rossmanEPANETUserManual2020} (see Fig. \ref{fig:CaseStudy} illustrating their layouts). The Three-node network consists of a reservoir, a pump, a junction, a pipe, and a tank, in addition to one booster station located at Junction J1 and one WQ sensor located at Tank TK1. Net1 has a reservoir, a pump, a tank, 9 junctions, and 12 pipes. Two booster stations are positioned at Junctions 1 and 6 of Net1, and two WQ sensors are placed at Junctions 4 and 9 within the network. The Richmond skeleton network is a schematic representation of the Richmond water distribution system, which is composed of one reservoir, 7 pumps, 41 junctions, 6 tanks, 8 valves, 37 pipes, 4 booster stations, and 3 WQ sensors.

\subsection{Hydraulic settings vs. water quality controllability}
\begin{figure}[t]
	\centering
	\includegraphics[width=0.49\textwidth]{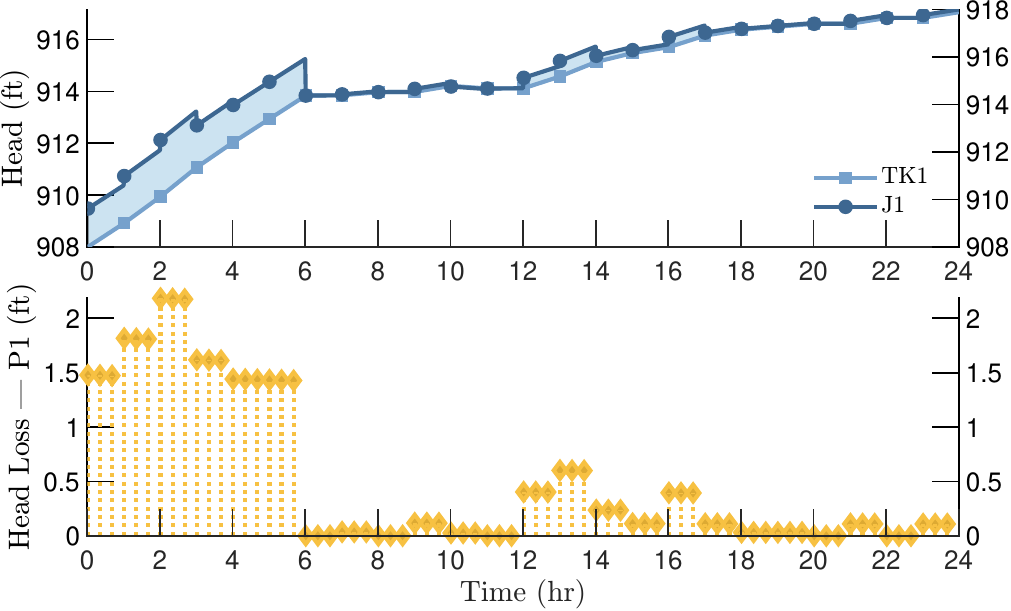}
	\vspace{-0.7cm}
	\caption{Head at Tank TK1 and Junction J1 (top) and the corresponding head loss in Pipe P1 (bottom) of the Three-node network.}\label{fig:HL_3N}
\end{figure}

Before we showcase the developed optimal pump scheduling approach, that is augmented with a desired level of controllability of water quality states, we first provide an investigation towards how varying the pump schedule actually affects the water quality dynamics. First, we run a hydraulic simulation on the Three-node network which has a simple layout where chlorine travels along a single available path. The velocity profile of P1 depicted in~Fig.~\ref{fig:HydvsWQContrl} is under the flow directions indicated in Fig. \ref{fig:CaseStudy}. The corresponding heads at TK1 and J1 and the head loss in P1 are shown in~Fig.~\ref{fig:HL_3N}. We construct the WQ-CG and employ the rank metric to assess the system's performance. It is important to note that in our analysis, we exclude Reservoir R1 and Pump M1 from the assessment as they are upstream of the booster station located at J1. This exclusion allows us to focus on evaluating the metric results specifically within the subspace of interest. We present these results as a percentage of the rank calculated within every hydraulic time-step, relative to the total number of states within this subspace ($n_\mathrm{J}+n_{s_1}+n_\mathrm{TK}$), number of segments $n_{s_1}$ of P1 is 100 segments. 

\begin{figure}[t]
	\vspace{-0.5cm}
	\centering
	\includegraphics[width=0.49\textwidth]{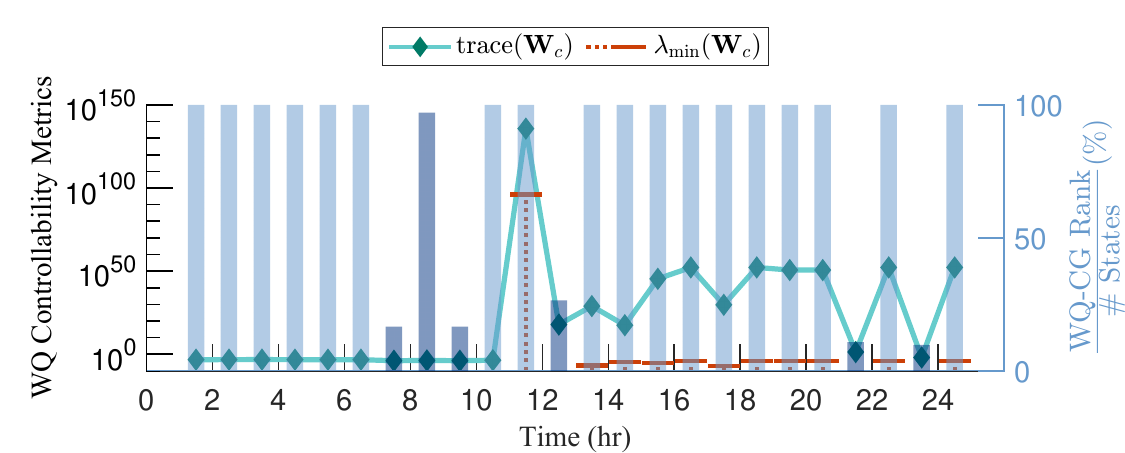}
	\vspace{-0.5cm}
	\caption{Controllability metrics, $\mr{trace}$ and $\lambda_{\min}$ of the controllable subspace Gramian $(\m{W}_{c})$ of the Three-node network vs. the percentage (\%) of the WQ-CG rank out of the \# states. Change in bars colors to highlight (darker shade) the windows where the WQ-CG is not full rank. }\label{fig:ContrlMetric_3N}
\end{figure}

In this particular scenario, the hydraulic time-step is set to 1 hour, while the WQ time-step is 10 seconds. Considering that Pipe P1 has a length of 1,000 ft, it is necessary for a water parcel to achieve a minimum velocity of 0.278 ft/sec in order to traverse the entire length of the pipe and reach Tank TK1 within the specified hydraulic time-step. This velocity directly influences the change of chlorine concentrations over time and space as expressed in the advection-reaction equation (Eq. \eqref{equ:PDE}). It is obvious in Fig. \ref{fig:HydvsWQContrl} that this characteristic has vital influence on the WQ controllability; when the velocities through P1 surpass this velocity boundary, the WQ-CG exhibits full rank, indicating full controllability of the system. On the other hand, in most cases when the velocities are lower, the system is uncontrollable with different deviations in the rank of the WQ-CG. It is worth mentioning that other factors are affecting the response of the system states to the inputs (in this case, the chlorine injections at J1). For instance, the rate of change of water volume at TK1 and the flows rates in the system's components, which are directly related to the head levels at the nodes and the head loss/gain at the links---refer to Fig. \ref{fig:HL_3N} and Appendix \ref{sec:App1}. Note that, the rank is calculated with keeping original MATLAB's machine epsilon which is equal to $2.2204\mathrm{e}{-16}$. Accordingly, with relative difference in that metric less than that machine epsilon, the two elements are considered dependent. 

In addition, in Fig. \ref{fig:ContrlMetric_3N} we showcase the $\mr{trace}(\m{W}_{c})$ and $\lambda_{\min}(\m{W}_{c})$ metrics of the controllable subspaces of the system over the same simulation period of 24 hours. It is noticeable that in many cases where the WQ-CG is full rank, there is change in the $\mr{trace}$ and $\lambda_{\min}$ values which reflects varying levels of energy stored in the system and their respective directions. Furthermore, the greater the disparity between the $\mr{trace}$ and $\lambda_{\min}$ values, the more pronounced the sparsity is in the reachability of WQ control. 
\begin{figure}[t]
	\centering
	\includegraphics[width=0.49\textwidth]{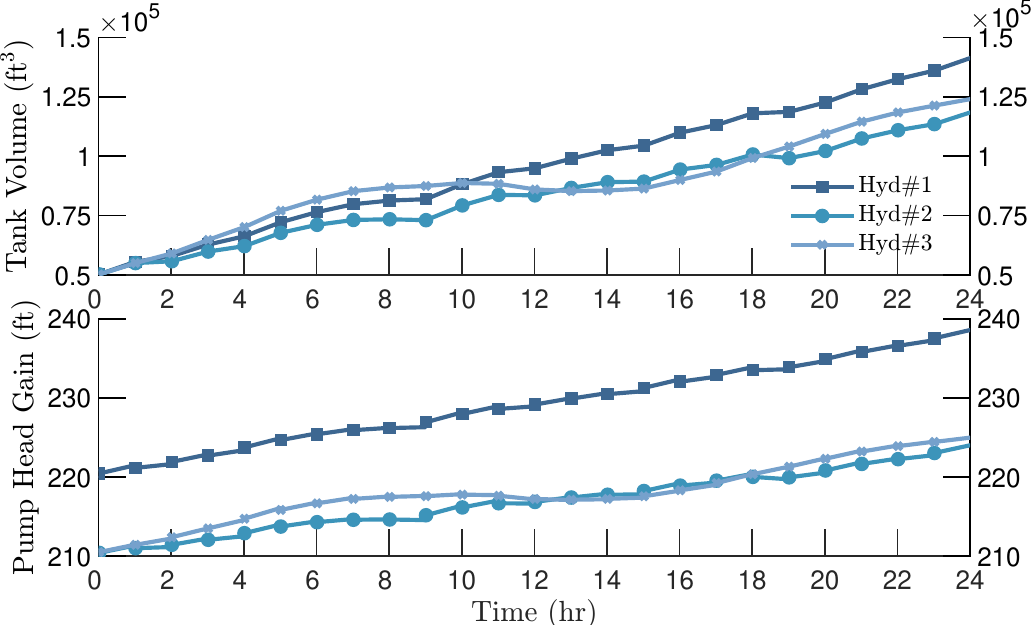}
	\vspace{-0.7cm}
	\caption{Change in the volume of Tank 11 and the operational pump head gain in Net1 under three different hydraulic scenarios; Hyd\#1, Hyd\#2, and Hyd\#3.}\label{fig:Net1_TankPump}
\end{figure}

We now apply three different hydraulic settings on Net1 resulting in the change in tank volume and the operational pump head gain depicted in Fig. \ref{fig:Net1_TankPump}. In the Hyd\#1 scenario, Tank 11 is filling throughout the whole simulation window. Whilst, Hyd\#2 and Hyd\#3 scenarios have smaller windows where Tank 11 is on demand. This results in different flow directions for specific pipes, which directly influence the water quality dynamics. In addition, it results in different total number of states (i.e, changes the number of segments into which each pipe is divided). All these scenarios are run over a 24 hours simulation period with a hydraulic time-step of 1 hour and a WQ time-step of 10 seconds. The total number of states for Hyd\#1 is 490, 470 for Hyd\#2, and 425 for Hyd\#3. 

\begin{figure*}[t]
	\centering	\subfloat[\label{fig:RankMetric_Net1}]{\includegraphics[keepaspectratio=true,width=0.49\textwidth]{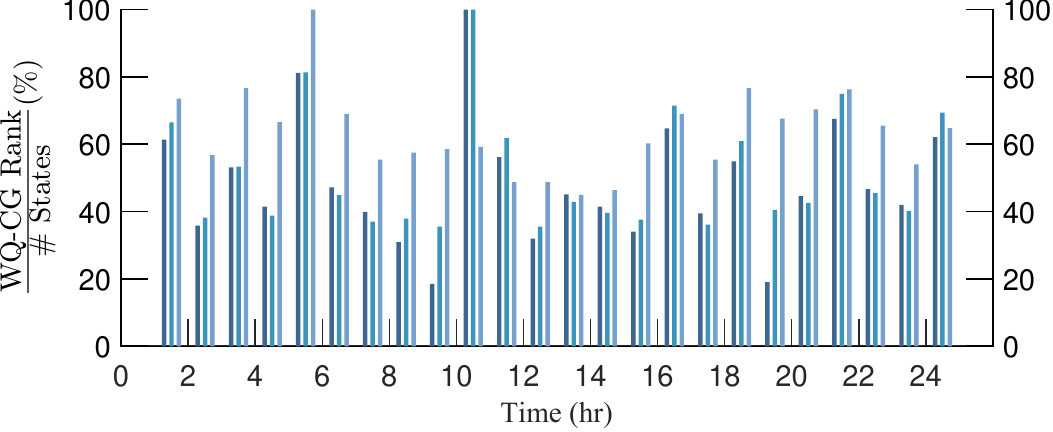}}{}\vspace{-0.1cm}
	\subfloat[\label{fig:TraceMetric_Net1}]{\includegraphics[keepaspectratio=true,width=0.45\textwidth]{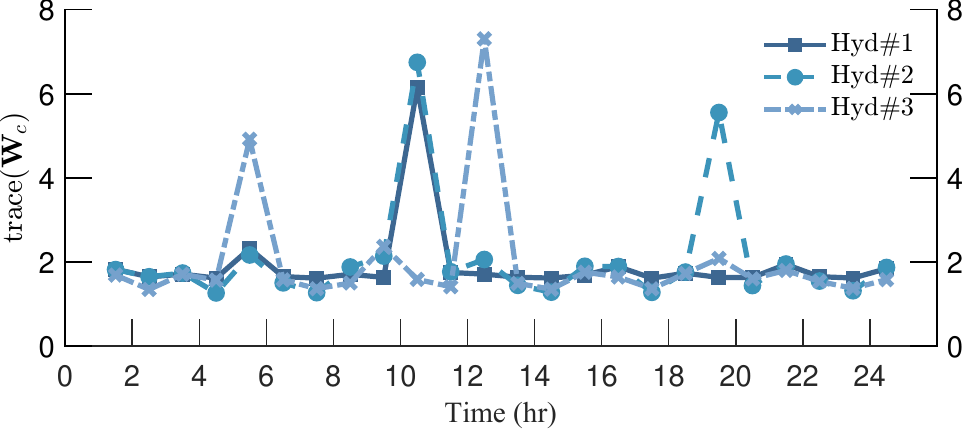}}{}\vspace{-0.1cm}
	\caption{For each of the three hydraulic scenarios (Hyd\#1, Hyd\#2, and Hyd\#3) applied on Net1, (a) the percentage (\%) of the WQ-CG rank out of the \# states and (b) $\mr{trace}$ of the controllable subspace Gramian.}\label{fig:ContrlMetric_Net1}
	\vspace{-0.5cm}
\end{figure*}

Fig. \ref{fig:RankMetric_Net1} shows the change in the rank of the controllability Gramian for each of the scenarios. The controllability Gramian does not reach full rank under the conditions where the booster stations are located at Junctions 1 and 6 and the water is drawn from Tank 11. In addition, the $\mr{trace}$ metric for the controllable subspace is calculated and the results are illustrated in Fig. \ref{fig:TraceMetric_Net1}. These values are affected by the direction in which the energy is distributed and the response of the system states to the inputs. Such response is affected by the input values in comparison to the systems states---the rate in which chlorine is injected into the system and the flow rates in the system. Note that, the results of the $\mr{log}\mr{det}$ metric for all hydraulic simulations under consideration are equal to the values of the $\mr{trace}$ presented in Fig. \ref{fig:TraceMetric_Net1}.
This observation implies that either of these metrics can be effectively employed in our study. In conclusion, each of the $\mr{rank}$, $\mr{trace}$, and $\mr{\lambda_{\min}}$ reflects an important behavior of the WQ dynamics and can be taken into consideration to reach the desired level of controllability over the system tailored to the scenario under focus.

\subsection{WQ controllability-aware optimal pump scheduling}
In this section we showcase the results of solving the decoupled and coupled pump optimal scheduling problems. The optimization problems are interfaced using YALMIP in MATLAB R2023a and solved using Gurobi and/or BMIBNB optimization solvers. The use of two optimization solvers is to compensate for the difference in the underlying problems formulated for each network and each scenario. The problems formulations differ in the their level of complexity and nonlinearity order. That being said, Gurobi is utilized to solve problems with lower nonlinearity order as it is a robust and fast solver capable of handling binary decision variables. Since Gurobi is limited under a highly nonlinear setting, the BMIBNB solver is used. BMIBNB is a global nonlinear solver capable of handling nonconvex problems, however as with all global solvers, the computational time is relatively more expensive in comparison to Gurobi. 

\begin{figure*}[t]
	\centering
	{\captionsetup[subfloat]{labelformat=empty}	\subfloat[\label{fig:3N_OptHyd1_TK1}]{\includegraphics[keepaspectratio=true,width=0.49\textwidth]{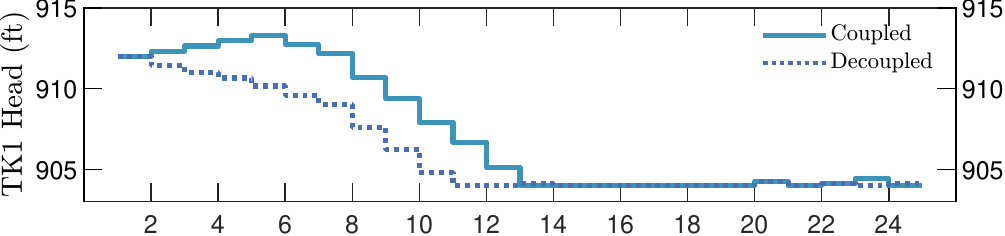}}{}\vspace{-0.7cm}} 
	{\captionsetup[subfloat]{labelformat=empty}	\subfloat[\label{fig:3N_OptHyd2_TK1}]{\includegraphics[keepaspectratio=true,width=0.49\textwidth]{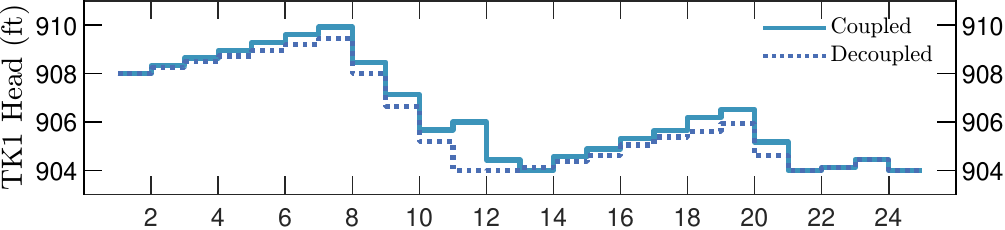}}{}} 
	{\captionsetup[subfloat]{labelformat=empty}	\subfloat[\label{fig:3N_OptHyd1_M1}]{\includegraphics[keepaspectratio=true,width=0.49\textwidth]{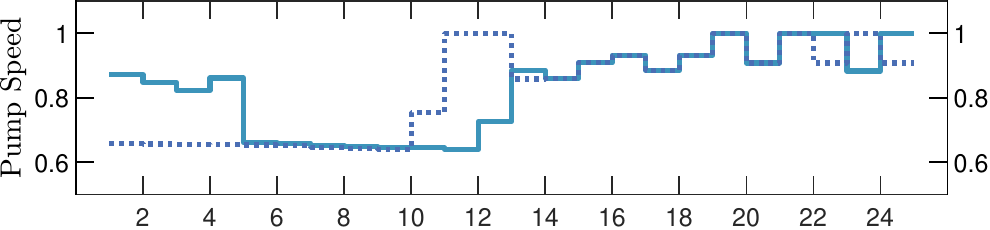}}{}\vspace{-0.7cm}} 
	{\captionsetup[subfloat]{labelformat=empty}	\subfloat[\label{fig:3N_OptHyd2_M1}]{\includegraphics[keepaspectratio=true,width=0.49\textwidth]{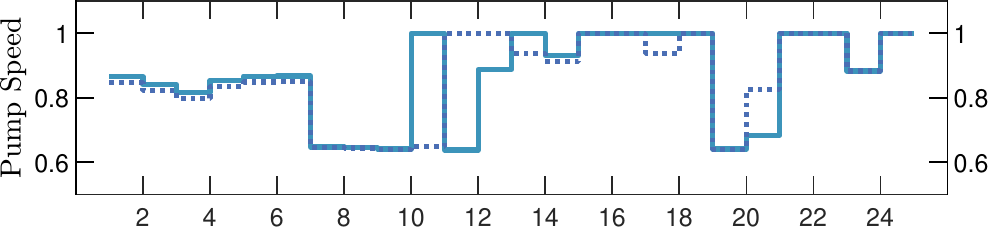}}{}} 
	\setcounter{subfigure}{0}
	\subfloat[\label{fig:3N_OptHyd1_PW1}]{\includegraphics[keepaspectratio=true,width=0.49\textwidth]{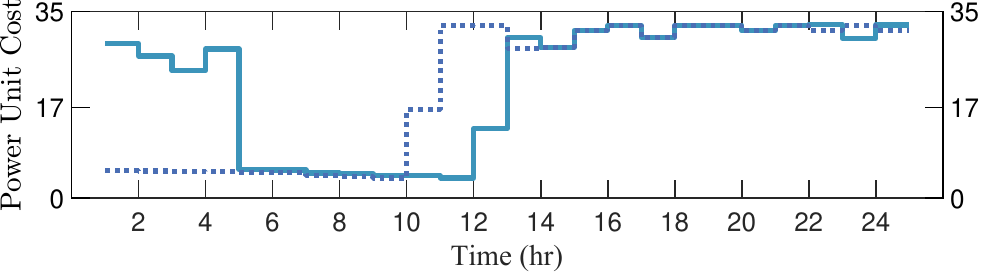}}{} 
	\setcounter{subfigure}{1}
	\subfloat[\label{fig:3N_OptHyd2_PW1}]{\includegraphics[keepaspectratio=true,width=0.49\textwidth]{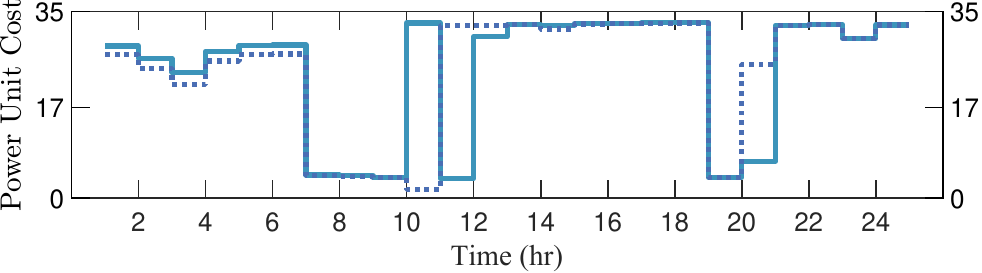}}{} \vspace{-0.15cm}
	\caption{Optimal Tank TK1 head, Pump M1 speed and power unit cost obtained by solving the decoupled and coupled pump scheduling problems of the Three-node network under the (a) first and (b) second hydraulic scenarios.}\label{fig:3N_OptHyd1&2}
	\vspace{-0.5cm}
\end{figure*}

With that in mind, first we apply the decoupled and coupled problem on the Three-node network under different scenario. The first scenario imposes no restrictions on the times when Tank TK1 is in an on-demand or off-demand state. Tank TK1 has a minimum of 904 ft, maximum of 924 ft, and an initial head level of 912 ft. Fig. \ref{fig:3N_OptHyd1_TK1} shows the optimal pump speed and Tank TK1 head levels obtained by solving the decoupled and coupled problems. For this scenario, the coupled problem is formulated to achieve higher and wider controllability at the beginning of the simulation period by constraining it with higher energy. This is done while filling the tank so that when it is in the on-demand state; it can supply the system with water that has sufficient chlorine levels. Filling the tank for this specific window is achieved by initially operating the pump at a higher speed. However, this increase in the pump speed is balanced later in the day when the tank supplies the network and the pump speed is lower compared to the decoupled problem. In this scenario, the WQ controller's performance is effectively improved by injecting chlorine at the start of the simulation period with fully controllable system. This approach reduces the reliance on the water volume stored in TK1, where chlorine concentration tends to decay over time. In addition, when utilizing the coupled optimization approach for pump scheduling, the total resulting cost is 8.9\% higher compared to the decoupled approach. In terms of computational efficiency, the solver completes the decoupled problem in 3.8 seconds, while the coupled problem requires 4.7 seconds to reach a solution. The second hydraulic scenario that we adopt has an initial tank head level at 908 ft and safe water level of 909.5 ft (Fig. \ref{fig:3N_OptHyd2_TK1}). The goal of maintaining a safe water level in TK1 leads to it being in the filling state at the beginning of the simulation period. Additionally, the coupled problem aims to higher controllability energy and coverage levels resulting in higher velocity in Pipe P1. In addition, the flow directions for both problems differ for the remainder of the simulation period, and the total operational cost of pump operation for both problems is comparable. In fact, the coupled problem's total cost is lower by 2\% in comparison to the decoupled problem. Moreover, under zero initial chlorine concentrations at Tank TK1 and Pipe P1, the WQ controller succeeds to achieve the set point concentration faster by 20 minutes.

\begin{figure}[h]
	\centering
	\subfloat[\label{fig:Net1_OptHyd1}]{\includegraphics[keepaspectratio=true,width=0.49\textwidth]{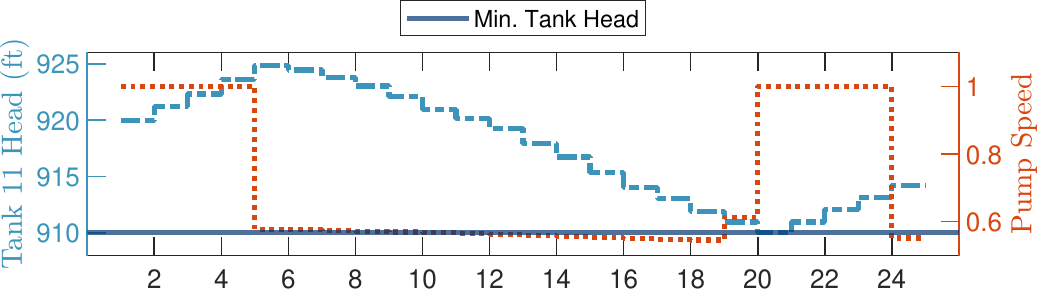}}{}\vspace{-0.3cm}
	\subfloat[\label{fig:Net2_OptHyd2}]{\includegraphics[keepaspectratio=true,width=0.49\textwidth]{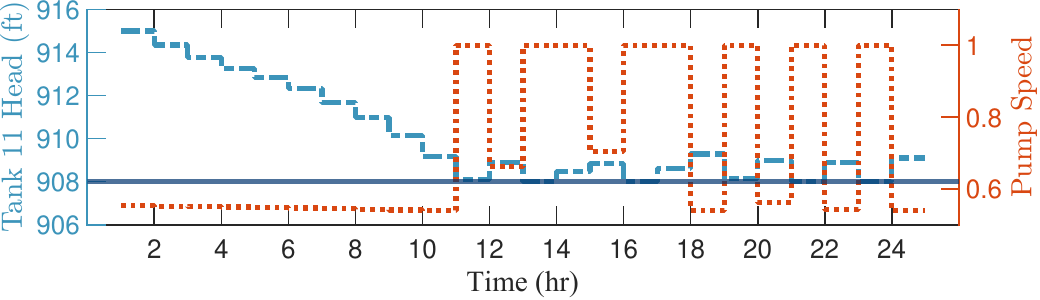}}{}\vspace{-0.3cm}
	\caption{Optimal Tank 11 head and Pump speed obtained by solving the coupled pump scheduling problem of the Net1 network under the (a) first and (b) second hydraulic scenarios.}\label{fig:Net1_OptHyd1&2}
\end{figure}

We then validate the performance of the coupled problem on Net1 network. It is worth noting that the pre-computation of the target controllability Gramian for this network requires approximately 2 seconds. Consequently, the runtime listed below exclusively pertains to solving the optimization problem separately for a 24-hour simulation period. As for the decoupled problem, the time required to solve it for this network using Gurobi is 8.7 sec and 161.1 sec using BMIBNB. On the other hand, the computational time for the coupled problem--where both solvers are used alternately during the hydraulic time-steps based on the constraints developed for each time-step--amounts to 145.2 seconds. For the first scenario, the initial Tank 11 head is 920 ft, while the minimum bound on the head is 910 ft. In this scenario, Junctions 2, 4, and 9 have water demands with different patterns and bases. Results from this simulation scenario are illustrated in Fig. \ref{fig:Net1_OptHyd1}. For the first 6 hrs of the simulation period, Tank 11 is being filled and accordingly is included in the target nodes while solving the rank-informed control problem. For the next simulation window till the 20th hour, Tank 11 is on demand allowing the pump to work on lower speeds. During this window, the coupled problem is formulated as an energy-driven problem with the network's junctions as the target nodes. However, the water head in Tank 11 reaches the minimum head level which requires activating a constraint in the optimization problem to recover the water level in the tank for the next simulation period. During the last 4 hrs of the simulation, the energy level required at start for some directions resulted in an infeasible solutions and accordingly, the problem is solved decoupled.

Another scenario for Net1, Tank 11 has initial head of 915 ft and minimum safety head of 908 ft. In addition, the stored water volume in the tank has sufficient chlorine levels of 1.5 mg/L to serve parts of the network, that is, it is excluded from the target nodes. Also, the demands' bases at the junctions are higher than the first scenario. As shown in Fig. \ref{fig:Net2_OptHyd2}, the pump speed is lower for the first half of the simulation period to fulfill the demands at the start of the network while Tank 11 is serving the rest. For the second half, the tank reaches the minimum safety head and keeps alternating between filling and emptying states. The coupled problem for this scenario is formulated to achieve higher controllability energy as the tank helps with serving the network with water that has sufficient chlorine levels and accordingly higher network chlorine coverage, specifically at the start of the simulation. The integration of the energy metrics in the hydraulic optimization problem for this scenario results in balancing the injections of the booster stations located at Junction 1 and Junction 6, in comparison to scenarios where the booster station at the start of the network is overworked and may run to an over capacity state. 

We now implement our proposed quality-aware pump control framework on the Richmond skeleton network which has a more branched network layout. This network comprises multiple elevated tanks at varying elevations, each serving distinct areas of the network with differing and varying water demands. For this network, we run multiple hydraulic scenario with varying target nodes throughout the simulation window. The selection of target nodes is based on the main pipeline leading to the tank during filling operations and the inclusion of dead-end points covered by booster stations. Essentially, the coupled problem for this network is designed to achieve extensive controllability coverage in specific network segments while maintaining a high controllability energy level within areas with low expected disinfectant residuals. This case study serves as a validation of the proposed framework's performance on a complex, branched real network and provides insights into the implementation of WQ control based on the obtained pump schedule. The average computational time to solve the coupled optimization problem for these scenarios is 432 seconds. Furthermore, across various scenarios involving distinct hydraulic settings and constraints, the WQ controller demonstrates improved and more efficient performance. This conclusion is drawn from the following observations made during the aforementioned scenarios: \textit{(i)} chlorine is injected by booster stations near tanks that are being filled to reach desired levels that are set to 2 mg/L; \textit{(ii)} chlorine injections are balanced among booster stations to distribute the workload. For instance, booster station at J9 is not overworked for the whole network and alternates based on tanks' on- and off-demand schedules and J15 supports the network while ensuring adequate residuals at J20; and \textit{(iii)} reduction in chlorine injections varies between 3-11\% when compared to the decoupled framework for scenarios with no firm restrictions on flow directions. However, for scenarios with such constraints, the increase in injections does not exceed 9\%, yet, dead-ends maintain sufficient chlorine residuals.

\section{Summary, Limitations, and Future Work}\label{sec:ConcRecom}
This paper provides a comprehensive analysis of how improving WQ controllability impacts the solution of the hydraulic settings optimization problem. This aims to enhance the performance of the WQ control and regulation algorithms. Specifically, these algorithms' performance is directly affected by the system hydraulics. To accomplish this, we incorporate the goal of improving various controllability metrics into the operational hydraulics optimization problem, aiming to achieve a targeted level of water quality controllability throughout the system. The performance of this approach is evaluated on three networks: a Three-node network, Net1, and the Richmond skeleton network. These networks vary in terms of scale and configurations. Additionally, different initial hydraulic and quality dynamics are examined for each network.

The results demonstrate the efficacy of the proposed approach in enhancing WQ controllability metrics, leading to a more efficient controller performance in achieving desired chlorine levels across the network. However, the enhancement of this performance is significantly influenced by variations in consumer demand patterns, leading to substantial shifts in system hydraulics in certain scenarios. Moreover, the network's configuration and topology has a pivotal role in determining the operational schedule and chlorine injections, with limitations imposed in some cases on the feasibility of chlorine reaching uncontrollable regions due to resultant flow directions. In conclusion, this approach is a step-forward in advancing the performance of regulating the water distribution networks dynamics and further improvement can be achieved by including these aspects while designing the network's layout and the functional constraints, which can be posed as work extensions in our group's future studies. Other future work encompasses the formulation of joint real-time WDNs control where hydraulics are updated with feedback that implies the foreseen effect on both; future hydraulic and WQ settings and dynamics not an optimization problem that is solved every hydraulic time-step with no feedback from the future of the impact of the control action.  

It is essential to recognize limitations in our study, particularly concerning the assumption of pre-allocated booster station locations throughout the network. Their locations directly impact water quality controllability. In our study, we set a specific configuration that ensures a satisfactory level of controllability across the entire network under different scenarios of hydraulic settings. Moreover, the chlorine control problem addressed in this paper does not account for the health risks associated with the formation of disinfectant by-products. To that end, we leave this problem for future work. This problem entails optimizing chlorine injections while ensuring that the formation of disinfectant by-products remains below standard levels, thereby mitigating potential health risks. 

\appendices
\section{WDN Hydraulic Modeling: Components}\label{sec:hyd_model}
We apply the principles of conservation of mass and energy to obtain the amount of water flowing in each network link and the head at each node. For all the network components, we give brief description of the equations that model these principles in the next section. 

For each of the following network elements, we model the hydraulics variables at/through this element depending on their characteristics and their connection to other elements.

\noindent -- \textit{Reservoirs:} We follow the valid assumption that reservoirs are infinite source of water with fixed head \cite{zamzamOptimalWaterPower2019}. Thence, the head at Reservoir $i$ is calculated as $h_i^\mathrm{R}(t+\Delta t_\mathrm{H})=h_i^\mathrm{R}(t)$.

\noindent -- \textit{Tanks:} We consider tanks with constant cross section along its height. Change in the head at the tank depends on its cross section's area and the algebraic difference between the inflows and outflows. The head at Tank $i$ is described as
\begin{equation}
	h_i^\mathrm{TK}(t+\Delta t_\mathrm{H})=h_i^\mathrm{TK}(t) + \frac{\Delta t_{\mathrm{H}}}{A_i^\mathrm{TK}} \Big(\sum_{j \in L_{\mathrm{in}}} {q_{\mathrm{in}}^{j}(t)} - \sum_{k \in L_{\mathrm{out}}} {q_{\mathrm{out}}^{k}(t)} \Big),
\end{equation}
where $A_i^\mathrm{TK}$ is the tank cross section's area; $j$ and $k$ are the counters for total $L_{\mathrm{in}}$ links flowing into the node and $L_{\mathrm{out}}$ links extracting flow from the node; and $q_{\mathrm{in}}^{j}(t)$ and $q_{\mathrm{out}}^{k}(t)$ are the inflows and outflows from these links connected to the node.

\noindent -- \textit{Junctions:} The conservation of mass law at junctions is expressed as
\begin{equation}
	\sum_{j \in L_{\mathrm{in}}} {q_{\mathrm{in}}^{j}(t)} - \sum_{k \in L_{\mathrm{out}}} {q_{\mathrm{out}}^{k}(t)} = q^{\mathrm{D}_\mathrm{J}}_i(t),
\end{equation} 
where $q^{\mathrm{D}_\mathrm{J}}_i(t)$ is the consumers' demand withdrawn from this junction.

\noindent -- \textit{Pipes:} As water flows in a pipe, it starts with the head of the upstream node and reaches the end-node head by applying the conservation of energy. The difference between the two heads corresponds to the difference in elevation between the upstream and the downstream node, as well as losses caused by friction along the pipe's length and localized minor losses (e.g., at bends, fittings, etc.). In our study, we neglect the minor losses as they are relatively small in comparison to the friction losses in water networks \cite{whiteFluidMechanics1966a}. That is, the change in head through Pipe $i$ connecting and flowing water between node $j$ and node $k$ is expressed in Eq. \eqref{eq:PipeLosses}---these nodes can be junctions, tanks, or reservoirs.
\begin{equation}\label{eq:PipeLosses}
	\Delta h^\mathrm{P}_i (t) = h_j(t) - h_k(t) = r_i q^\mathrm{P}_i(t) |q^\mathrm{P}_i(t)|^{\mu-1},
\end{equation}
where $q^\mathrm{P}_i(t)$ is the pipe flow; $r_i$ is the pipe resistance coefficient, which is a function of pipe size, length, and material; and $\mu$ is the constant flow exponent. These parameters' values depend on the chosen head loss formula. In our study we use the Hazen-Williams equation; $\mu=1.852$ \cite{linsleyWaterResourcesEngineering1997}.

\noindent -- \textit{Pumps:} As an active component with variable speeds, pumps can provide the system with different values of head gain according to its operating speed and the corresponding head-flow relationship \cite{linsleyWaterResourcesEngineering1997}. For Pump $i$ adding energy to water flowing from node $j$ to node $k$, the head gain is calculated as
\begin{equation}\label{eq:PumpLosses}
	\Delta h^\mathrm{M}_i (t) = h_j(t) - h_k(t) = -s^2_i(t) (h_i^0-\alpha_i (s^{-1}_i(t) q^\mathrm{M}_i(t))^{\nu_i}),
\end{equation}
where $s_i(t)$ is the pump relative speed varying between 0 and the maximum speed $s_i^{\max}$, which is a positive unique value ($s_i^{\max}>0$) for each pump that depends on its characteristics and impeller size; $h_i^0$ is the shutoff head; $q^\mathrm{M}_i(t)$ is the pump flow; and $\alpha_i$ and $\nu_i$ are pump characteristics coefficients. Note that the head gain is strictly negative, as the pump provides the water with more energy, causing the head at the delivery node to exceed that at the suction node, and no back-flow is allowed. 

\noindent -- \textit{Valves:} In our model, we consider pumps to be the only controller of the system. Therefore, we formulate and solve pump operation problem to obtain the optimal pumping schedule that fulfills the water flow and head constraints. That being the case, valves in our networks are considered as \textit{on-off valves} and they have two states; fully closed or fully open. The knowledge of valve state is predetermined along the simulation period. In the case of fully closed, the two nodes connected by the valve are considered decoupled. For the other case of fully open valve, it is treated as a straight pipe section with minor losses~\cite{rennelsPipeFlowPractical2022} that can be expressed as in Eq. \eqref{eq:ValveLosses} for Valve $i$ connecting nodes $j$ and $k$.
\begin{equation}\label{eq:ValveLosses}
	\Delta h^\mathrm{V}_i(t) = h_j(t) - h_k(t) = m_i q^\mathrm{V}_i(t) |q^\mathrm{V}_i(t)|,
\end{equation}
where $q^\mathrm{V}_i(t)$ is the flow through the valve and $m_i$ is the minor losses coefficient that depends on the valve type (e.g., ball, butterfly, gate, etc.) and its cross-sectional area.

\section{WQ modeling in WDNs}\label{sec:WQModel}
\subsection{Conservation of chlorine mass in nodes}\label{sec:ConvMass}
For reservoirs, tanks, and junctions, the principles of conservation of mass are applied. Reservoirs are considered a continuous source of chlorine with constant concentrations over time; for a Reservoir $i$, $c^\mathrm{R}_i(t+\Delta t_{\mathrm{WQ}})=c^\mathrm{R}_i(t)$. On the other hand, junctions and tanks are assumed to have complete immediate mixing at place with no storage for junction and changing volume with time for tanks\cite{boulosComprehensiveWaterDistribution2006,burkhardtModelingFateTransport2017}. Accordingly, if node $i$ is defined as a Junction, chlorine concentration at this node is calculated as
\begin{equation}\label{equ:mb-junc} 
	c_i^\mathrm{J}(t)= \frac{\sum_{j \in L_{\mathrm{in}}} \textcolor{violet}{q_{\mathrm{in}}^{j}(t)} c_\mathrm{in}^j(t)+\textcolor{black}{q^\mathrm{B}_i(t)} c^\mathrm{B}_i(t)}{\textcolor{violet}{q^{\mathrm{D}_\mathrm{J}}_i(t)}+\sum_{k \in L_{\mathrm{out}}} \textcolor{violet}{q_{\mathrm{out}}^{k}(t)}}.
\end{equation}
While, if it is defined as a Tank, $c_i^\mathrm{TK}$ is calculated as follows
\begin{equation}\label{equ:tank2}
	\small	\begin{split}
		& \textcolor{violet}{V_i^\mathrm{TK}(t + \Delta t_{\mathrm{WQ}})} c_i^\mathrm{TK}(t+ \Delta t_{\mathrm{WQ}})= \textcolor{violet}{V_i^\mathrm{TK}(t)} c_i^\mathrm{TK}(t) 
		\\ & +\sum_{j \in L_{\mathrm{in}}} \textcolor{violet}{q^j_\mathrm{in}(t)} c^j_\mathrm{in}(t) \Delta t_{\mathrm{WQ}}
		+\textcolor{black}{V^\mathrm{B}_i(t+\Delta t_{\mathrm{WQ}})} c^\mathrm{B}_i(t+\Delta t_{\mathrm{WQ}})\\
		&\hspace{-0.4cm} - \sum_{k \in L_{\mathrm{out}}} \textcolor{violet}{q^k_\mathrm{out}(t)} c_i^\mathrm{TK}(t) \Delta t_{\mathrm{WQ}}
		+R^\mathrm{TK}(c_i^\mathrm{TK}(t)) \textcolor{violet}{V_i^\mathrm{TK}(t)} \Delta t_{\mathrm{WQ}},
	\end{split}
\end{equation}
where $c_\mathrm{in}^j(t)$ is the concentration in the inflow solute; $V_i^\mathrm{TK}(t)$ is the water volume of the tank, i.e., $V_i^\mathrm{TK}(t)=A_i^\mathrm{TK} h_i^\mathrm{TK}(t)$; $q^\mathrm{B}_i(t)$ and $V^\mathrm{B}_i(t+\Delta t)$ are the flow and the volume of chlorine injected to the node with concentration $c^\mathrm{B}_i(t)$ by booster station if located; and $R^\mathrm{TK}(c^\mathrm{TK}_i(t))$ is the decay and reaction expression in tanks (refer to Appendix~\ref{sec:SSmodel}). Booster stations located at tanks offer water utility operators the ability to maintain constant chlorine concentrations at outflow pipes. This scenario can be accommodated by incorporating these constraints into the control problem. In our paper, we present a generalized model that can be customized to address this scenario based on the network being studied, as well as other scenarios with changing desired levels of chlorine to be maintained.

Booster stations located at tanks can be utilized by water utilities operators to attain constant chlorine concentrations at outflows pipes. This scenario is achievable by integrating these constraints in the control problem. In our paper, we consider a generalized model that can be tailored to this scenario or not according to the network under study. It is worth mentioning that from an operational perspective, water operators tend to inject chlorine dosages in tanks to maintain constant chlorine concentrations in th outlet pipe 

\subsection{Chlorine transport and reaction model in links}\label{sec:TranspReact}
In the water quality model, pumps and valves are considered links with no actual length and accordingly and accordingly there are no changes in the chemical concentration from the upstream node. That is, for Pump $i$ and Valve $k$ place after Junction $j$, concentrations are expressed as $c_i^{\mathrm{M}}(t+\Delta t_{\mathrm{WQ}}) = c_j^{\mathrm{J}}(t+\Delta t_{\mathrm{WQ}}),$ and $c_k^{\mathrm{V}}(t+\Delta t_{\mathrm{WQ}}) = c_j^{\mathrm{J}}(t+\Delta t_{\mathrm{WQ}}).$

Nonetheless, the transport and reaction model in pipes is simulated by the one-dimensional advection-reaction (1-D AR) partial differential equation (PDE), which for Pipe $i$ is expressed as
\begin{equation}\label{equ:PDE}
	\partial_t c_i^\mathrm{P}=-\textcolor{violet}{v_i(t)} \partial_x c_i^\mathrm{P} + R^\mathrm{P}(c_i^\mathrm{P}(x,t)),
\end{equation}
where $c^\mathrm{P}_i(x,t)$ is concentration in pipe at location $x$ along its length and time $t$; $v_i(t)$ is the mean flow velocity which is a hydraulic variable that characterizes the rate at which water flows through the pipe; and $R^\mathrm{P}(c^\mathrm{P}_i(x,t))$ is the decay reaction expression in pipes (more explanation is given in Appendix~\ref{sec:SSmodel}). Eq. \eqref{equ:PDE} is discretized over a fixed spatio-tamporal grid using the \textit{Explicit Upwind scheme}---Eulerian Finite-Difference based method \cite{korenRobustUpwindDiscretization1993,elsherifControltheoreticModelingMultispecies2023}. This scheme is conditionally stable by satisfying the Courant-Friedrichs-Lewy condition (CFL). This condition puts limits on the Courant number $\textcolor{violet}{{\lambda}_i(t)}=\frac{\textcolor{violet}{v_i(t)} \Delta t}{\Delta x_i}$ to be $0<\textcolor{violet}{{\lambda}_i(t)} \leq 1$, for a Pipe $i$.

Subsequently, the Pipe $i$ with length $L_{i}$ is split into a number of segments $n_{s_i}=\Big\lfloor \frac{L_i}{\textcolor{violet}{v_i(t)} \Delta t} \Big\rfloor$ of length $\Delta x_i= \frac{L_i}{n_{s_i}}$, note that the symbol $\left \lfloor \cdot \right \rfloor $ donates the floor function, which takes a real number as input and returns the greatest integer less than or equal to that number. The chemical concentrations for the pipe segments, ranging from the first segment $c^\mathrm{P}i(1,t+ \Delta t_{\mathrm{WQ}})$ to all segments in between along the pipe's length, and reaching the last segment $c^\mathrm{P}i(s,t+ \Delta t_{\mathrm{WQ}})$, are calculated as expressed in \eqref{eq:PipeConcDisc}. This calculation assumes that Junction $j$ is upstream of this pipe. 
\begin{multline}\hspace{-0.1cm}\label{eq:PipeConcDisc}
	\hspace{-0.25cm}
	\begin{aligned}
		& \begin{bmatrix}
			c^\mathrm{P}_i(1,t+ \Delta t_{\mathrm{WQ}}) \\ c^\mathrm{P}_i(2,t+ \Delta t_{\mathrm{WQ}}) \\ \vdots \\ c^\mathrm{P}_i(s-1,t+ \Delta t_{\mathrm{WQ}}) \\ c^\mathrm{P}_i(s,t+ \Delta t_{\mathrm{WQ}}) 
		\end{bmatrix} = (1-\textcolor{violet}{{\lambda}_i(t)}) \begin{bmatrix} c^\mathrm{P}_i(1,t) \\ c^\mathrm{P}_i(2,t) \\ \vdots \\ c^\mathrm{P}_i(s-1,t) \\ c^\mathrm{P}_i(s,t) \end{bmatrix} \\ & +\textcolor{violet}{{\lambda}_i(t)} \begin{bmatrix} c^\mathrm{J}_j(t) \\ c^\mathrm{P}_i(1,t) \\ \vdots \\ c^\mathrm{P}_i(s-2,t) \\ c^\mathrm{P}_i(s-1,t) \end{bmatrix} + \Delta t_{\mathrm{WQ}} \begin{bmatrix} R^\mathrm{P}(c^\mathrm{P}_i(1,t)) \\ R^\mathrm{P}(c^\mathrm{P}_i(2,t)) \\ \vdots \\ R^\mathrm{P}(c^\mathrm{P}_i(s-1,t)) \\ R^\mathrm{P}(c^\mathrm{P}_i(s,t)) \end{bmatrix}.
	\end{aligned}
\end{multline}

\subsection{Single-species decay model}\label{sec:SSmodel}
The single-species decay model is a first-order model where the chlorine concentrations are decaying due to wall reaction dynamics in pipes and bulk reaction dynamics in both; pipes and tanks. Henceforward, the chlorine decay reaction rates for Pipe $i$ and Tank $j$ are $k_i^\mathrm{P} = k_{b}+\frac{2k_{w}k_{f}}{r_{\mathrm{P}_i}(k_{w}+k_{f})},\,\,\,\,\, k_j^\mathrm{TK} = k_{b}$, where $ k_{b}$ is the bulk reaction rate constant; $k_{w}$ is the wall reaction rate constant; $k_{f}$ is the mass transfer coefficient between the bulk flow and the pipe wall; and $r_{\mathrm{P}_i}$ is the pipe radius.  It is noteworthy that these parameters are influenced by many factors, which vary between water chemistry and contact time for bulk parameters, and pipe material, pipe age, and biofilm growth for the wall decay parameters. For a more comprehensive understanding of how these parameters are determined and the factors that impact them, we refer readers to~\cite{fisher2011evaluation,fisher2017comprehensive,fisher2017new,fisherFrameworkOptimizingChlorine2018}. Eventually, the decay and reaction expressions for Segment $s$ of Pipe $i$ and Tank $j$ are
\begin{equation}
	R^\mathrm{P}(c^\mathrm{P}_i(s,t)) = - k_i^\mathrm{P} c^\mathrm{P}_i(s,t), \;\; R^\mathrm{TK}(c_i^\mathrm{TK}(t)) = - k_i^\mathrm{TK} c^\mathrm{TK}_i(t).
\end{equation}

\section{Controllable subspace}\label{sec:reach_control}
The reachable subspace includes all the states that a system can reach over time, without necessarily applying any specific control inputs. However, for a controllable subspace, these states can be reached by applying specific control inputs.

For an uncontrollable WQ system with total number of states $n_x$, let the rank of the controllability matrix/Gramian be $k < n_x$. Then there exists a nonsingular matrix $\mT \in \mathbb{R}^{n_x \times n_x}$ such that
\begin{equation}\label{eq:UncontrlSysDecomp}
	\begin{split}
		& \bar{\mA}_\mathrm{WQ} =  \mT \mA_\mathrm{WQ} \mT^{-1} = \begin{bmatrix}
			\bar{\mA}_{\mathrm{WQ},11} & \bar{\mA}_{\mathrm{WQ},12} \\
			\boldsymbol{0} & \bar{\mA}_{\mathrm{WQ},22}
		\end{bmatrix}, \\
		& \bar{\mB}_\mathrm{WQ} = \mT \mB_\mathrm{WQ} = \begin{bmatrix}
			{\bar{\mB}_{\mathrm{WQ},1}} \\ \boldsymbol{0}
		\end{bmatrix},
	\end{split}
\end{equation}
where $\bar{\mA}_{\mathrm{WQ},11}, \; \bar{\mA}_{\mathrm{WQ},12}$ and $\bar{\mA}_{\mathrm{WQ},22}$ have dimensions of $k \times k, \; k \times (n_x-k)$ and $(n_x-k) \times (n_x-k)$, and $\bar{\mB}_{\mathrm{WQ},1}$ has $k$ rows. Namely, $\bar{\mA}_{\mathrm{WQ},11}$ and $\bar{\mB}_{\mathrm{WQ},1}$ defines a controllable subspace. Readers are referred to~\cite{dattaNumericalMethodsLinear2004} for the theorem pertaining to the development of reachable subspace decomposition.

\section{Decoupled pump scheduling Formulation}\label{sec:de_pump_schedule}
We first follow the approximation approach for the system components proposed in \cite{menkeApproximationSystemComponents2015} to formulate the pump control problem. The system components include the head loss in pipes, head loss in valves, head gain in pumps, and pump power consumption. For the head losses in pipes and valves, we apply piecewise linear approximations that transform \eqref{eq:PipeLosses} and \eqref{eq:ValveLosses} to multiple linear constraints.

\begin{figure*}[t!]
	\centering
	\subfloat[\label{fig:PipeCurve}]{\includegraphics[keepaspectratio=true,scale=0.3]{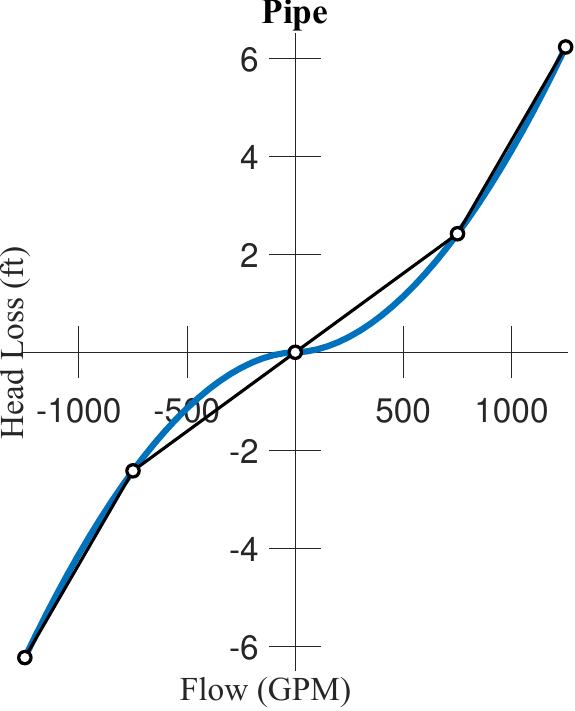}}{} \hspace{1.4cm}
	\subfloat[\label{fig:PumpCurve}]{\includegraphics[keepaspectratio=true,scale=0.3]{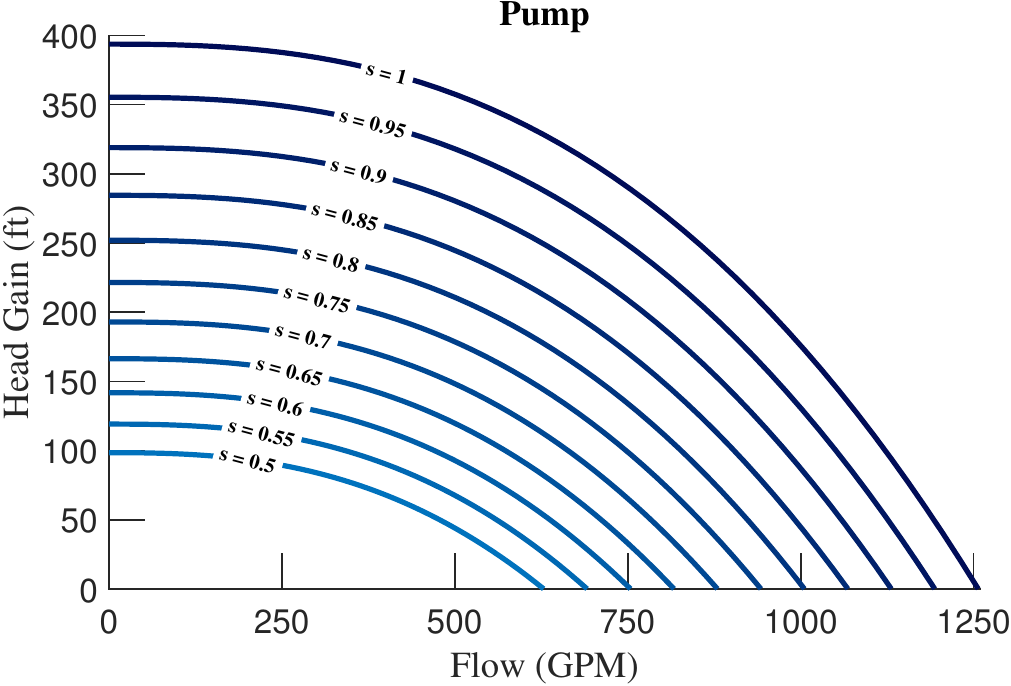}}{}\hspace{1.2cm}
	\subfloat[\label{fig:PumpPower}]{\includegraphics[keepaspectratio=true,scale=0.3]{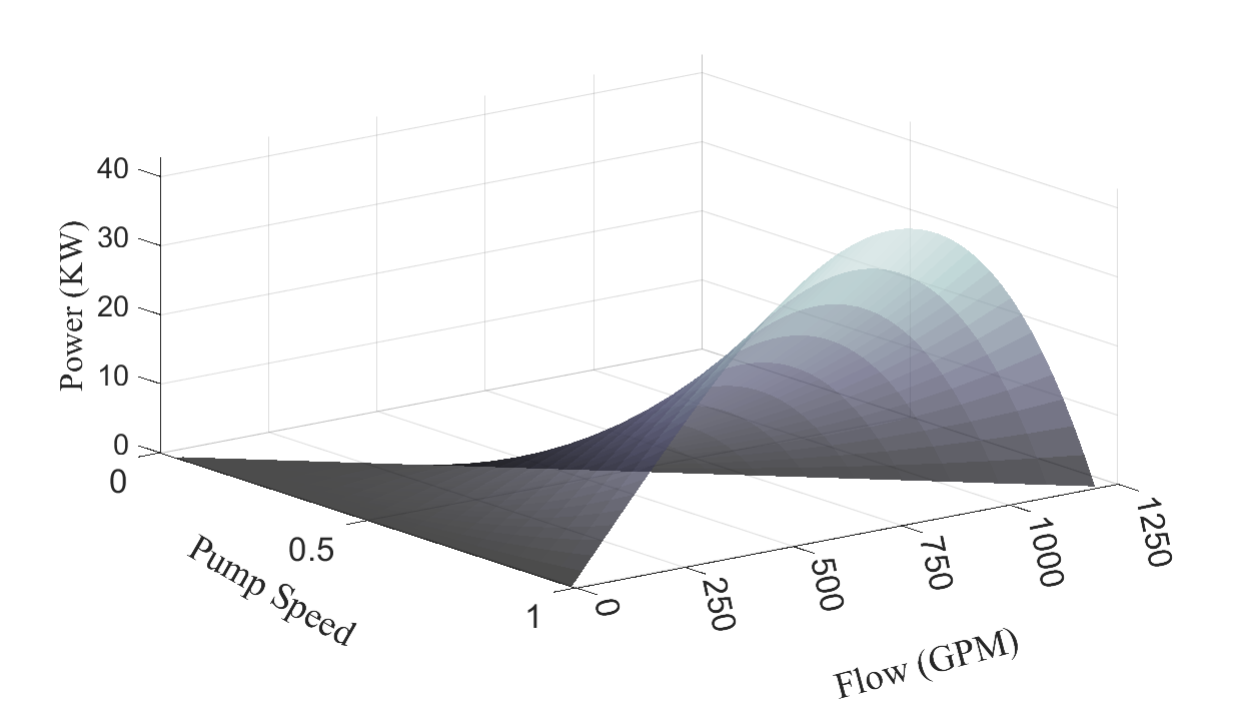}}{}
	\vspace{-0.1cm}
	\caption{(a) Linear, (b) variable-speed pump curve, and (c) the resultant power consumption ($ h_0 = 393.7, \alpha = 3.7 \times 10^{-6}, \nu = 2.59$). 
	}
	\label{fig:PipePumpCurve}
	\vspace{-0.5cm}
\end{figure*}
\setlength{\textfloatsep}{0pt}

Specifically, each pipe/valve's head loss curve is segmented into linear segments, which are determined by connecting points that can be calculated offline as a pre-optimization step. For the curve and piecewise linearization presented in Fig. \ref{fig:PipeCurve} as an example, four points are located and the segmented lines are connected. Using the equations for these lines, three constraints are added to the optimization problem for this specific pipe. When aiming for closer fits (i.e., employing more segmentation), a drawback emerges in the context of large network models, which is that scalability is negatively affected. For formulations of $N_\mathrm{PW}$ pieces, the constraints for Pipe $i$ connecting and flowing water between node $j$ and node $k$, are as follows
\begin{subequations}\label{equ:PipeConstraints}
	\begin{align}
		&\hspace{-0.3cm}	h_j(t) - h_k(t) - \sum_{n=1}^{N_\mathrm{PW}} \tilde{m}_n  \zeta_n(t) - \sum_{n=1}^{N_\mathrm{PW}} \tilde{b} \omega_n(t)  = 0,\label{equ:PipeHeadLossLin} \\
		&	q^\mathrm{P}_i(t) - \sum_{n=1}^{N_\mathrm{PW}} \zeta_n(t) =0, \label{equ:PipeFlowEq} \\
		&	\sum_{n=1}^{N_\mathrm{PW}} \omega_n(t)  = 1, \label{equ:PipeSegmentSelect} \\
		&  \hspace{-0.35cm}\begin{cases} 
			-\zeta_n(t) + q_{n,\min} \omega_n (t) \leq 0, \\
			\zeta_n(t) - q_{n,\max} \omega_n (t) \leq 0, 
		\end{cases}\label{equ:PipeBoundSeg}
	\end{align}
\end{subequations} 
where $\tilde{m}_n$ and $\tilde{b}$ are the $n$ segment's line slope and intercept, while $q_{n,\min}$ and $q_{n,\max}$ are the flow boundary limits. In addition, for the same segment $n$, $\zeta_n(t)$ and the binary $\omega_n(t)$ are decision variables to enable falling within the right segment range. Constraint \eqref{equ:PipeHeadLossLin} represents the linearized head loss through the pipe within this segment, \eqref{equ:PipeFlowEq} defines the pipe segments flow equality constraint, \eqref{equ:PipeSegmentSelect} allows the segment selection, and \eqref{equ:PipeBoundSeg} are the boundary constraints for each segment. By adding these equality and inequality constraints to the optimization problem, new integer decision variables are introduced; \textit{mixed integer programming}. 

Moreover, in contrast to study \cite{menkeApproximationSystemComponents2015} we use variable speed pumps instead of fixed  speed pumps, thereby making the problem more general yet different. This difference arises due to the variations in pump curves based on the selected pump speed; see Fig. \ref{fig:PumpCurve}. Therefore, we follow the proposed methodology by \cite{menkeModelingVariableSpeed2016,horvathConvexModelingPumps2019}, making minor modifications to transform the decision variables into the flow through the pump and pump speed, all while ensuring convexity. This transformation is achieved by applying the affinity rules that govern the relationship between pump shaft speed, discharge, and head gain. These rules relate the discharge and head gain to the shaft speed via parabolic relations. Specifically, pump discharge exhibits a linear relationship with pump speed, whereas head gain depends on the square of the speed. As a pre-optimization step, both the pump performance curve and the corresponding power consumption that we aim to minimize are approximated to formulate two convex expressions to be integrated in the optimization problem. 

First, we approximate the characteristic curve of Pump $i$ from Eq. \eqref{eq:PumpLosses} to the following
\begin{equation}\label{equ:AppPumpCurve}
	\Delta h_i^\mathrm{M}(t)\big|_{\mathrm{App}} = \beta_1 \Big(q^\mathrm{M}_i(t)\Big)^2 + \beta_2 q^\mathrm{M}_i(t) + \beta_3 \Big(s^\mathrm{M}_i(t)\Big)^2 + \beta_4,
\end{equation}
where $\beta_1, \beta_2, \beta_3$ and $\beta_4$ are coefficient calculated by minimizing the error between $\Delta h_i^\mathrm{M}$ in Eq. \eqref{eq:PumpLosses} and Eq. \eqref{equ:AppPumpCurve} with $\beta_1, \beta_3 \geq 0$ to ensure convexity.

Furthermore, alongside the system dynamics and their approximations addressed in the preceding sections of the paper, we also account for the physical constraints on the head levels and flow values among the various network components---expressed in Eq. \eqref{equ:HydPhysConstraints}.  Pump speed vector $\vs(t)$ is constrained to be between $\boldsymbol{0}$ and $\vs_{\max}$, where the zero values indicate that the pump is off. These considerations can all be written as box constraints
\begin{equation}\label{equ:HydPhysConstraints}
	\begin{split}
		\vw(t) \in [\vw_{\min}, &   \vw_{\max}], \;\;
		\vl(t) \in [\vl_{\min},    \vl_{\max}], \\
		\vz(t) \in [\vz_{\min}, &   \vz_{\max}], \;\;
		\vs(t) \in [\boldsymbol{0}, \vs_{\max}].
	\end{split}
\end{equation}

Lastly, the objective function for this problem enforces minimizing the cost of power consumption by pumps. This objective function is expressed as
\begin{equation}\label{equ:PumpPowerActual}
	\Pi(t)= \varphi_\mathrm{EL} \sum_i^{n_\mathrm{M}} \frac{\rho_\mathrm{W}g}{\eta_i(t)} \Delta h_i^{\mathrm{M}}(t) q_i^{\mathrm{M}}(t),
\end{equation}
where $\varphi_\mathrm{EL}$ is the price of each kW of electricity per hour (\$/kWh), $\rho_\mathrm{W}$ is water density, $g$ is the gravity acceleration, $\eta_i(t)$ is the efficiency of Pump $i$ under a head gain of $\Delta h_i^{\mathrm{M}}(t)$ and flow of $q_i^{\mathrm{M}}(t)$.

As illustrated in Fig. \ref{fig:PumpPower}, the power consumption of pumps exhibits a nonlinear relationship with respect to the flow rate, and this relationship shifts when the pump speed is altered. For a specific pump speed, this objective function can be approximated to a convex second-order function \cite{horvathConvexModelingPumps2019}. However, when the actual pump speed deviates from the speed at which the curve is approximated, the optimized pump power consumption varies substantially from the actual power consumption---a limitation of \cite{horvathConvexModelingPumps2019} to work in a vacancy of the approximation region. To overcome this issue, we approximate the objective function to be function in the pump speed and flow. This approach allows us to avoid constructing an approximate formulation reliant on head gain with inherent error, thereby preventing an increase in inaccuracies. We define this function for Pump $i$ to be
\begin{equation}\label{equ:PumpPowerApp}
	\begin{split}
		\Pi_{\mathrm{App}}(t) &= \theta_1+\theta_2 q^\mathrm{M}_i(t)+\theta_3 \Big(q^\mathrm{M}_i(t)\Big)^2 +\theta_4 s^\mathrm{M}_i(t) \\ 
		& +\theta_5 \Big(s^\mathrm{M}_i(t)\Big)^2 +\theta_6 q^\mathrm{M}_i(t) s^\mathrm{M}_i(t).
	\end{split} 
\end{equation}
The vector $\m{\theta}:=\{\theta_i | i \in \{1, \cdots, 6\}\}$ collects the coefficients of the approximate second-order power consumption function. These coefficients are derived by solving a straightforward optimization problem aimed at ensuring the convexity of this function. This function is convex under the condition that its Hessian is positive semidefinite. The Hessian is defined to be $\mathbfcal{H}=\begin{bmatrix}
	2 \theta_3 & \theta_6 \\
	\theta_6 & 2 \theta_5
\end{bmatrix}$. That is, we obtain these coefficients by solving the following optimization problem \eqref{equ:PumpPowerCoeffOptProb} after pre-calculating the power consumption \eqref{equ:PumpPowerActual} for different $N_{\mathrm{M,OP}}$ operating points in the domain of the characteristic curve of Pump $i$. Thereby, $q_{i,j}^M$ and $s_{i,j}^M$ represent the flow rate and relative speed, respectively, at the $j$-th operating point ($j = 1, \ldots, N_{\mathrm{M,OP}}$) on the characteristic curve of Pump $i$.
\begin{subequations}\label{equ:PumpPowerCoeffOptProb}
	\begin{align}
		\hspace{-0.75cm}	\underset{\m{\theta}}{\mbox{minimize}} \;\; & \sum_{j=1}^{N_{\mathrm{M,OP}}} (\Pi_{\mathrm{App}}(q^\mathrm{M}_{i,j},s^\mathrm{M}_{i,j})-\Pi(q^\mathrm{M}_{i,j},s^\mathrm{M}_{i,j}))^2 \\
		\begin{split}\label{eq:DecHydOptConst_2}
			\mbox{subject to}\;\;& \mathbfcal{H} \succeq 0.
		\end{split}	
	\end{align}		
\end{subequations}

\section{Variables and Units}\label{sec:App2}
In this appendix, we present the variables used in this paper in Tab. \ref{tab:units}, along with their respective unit dimensions, and the units employed for each variable in the case studies (Section \ref{sec:CaseStudies}).
\begin{table}[h!]
	\scriptsize
	\renewcommand{\arraystretch}{1.1}
	\setlength{\tabcolsep}{3pt} 
	\centering
	\caption{Variables and Units}
	\vspace{-0.3cm}
	\label{tab:units}
	\resizebox{\linewidth}{!}{%
	\begin{tabular}{@{}llll@{}}
		\toprule
		Variable & Description & Dimensions & Units \\
		\midrule
		$t$ & Time & $[T]$ & second, minute, hour \\
		$\Delta t_\mathrm{WQ}$ & Water quality time-step & $[T]$ & second, minute, hour  \\
		$\Delta t_\mathrm{H}$ & Hydraulic time-step & $[T]$ & second, minute, hour \\
		$h$ & Head  & $[L]$ & feet \\
		$A$ & Area & $[L^2]$ & square feet ($\text{ft}^2$)  \\
		$q$ & Flow rate & $[L^3T^{-1}]$ & gallon per minute (GPM) \\
		$s$ & Pump relative speed & --- & --- \\
		$c$ & Chemical concentration & $[ML^{-3}]$ & milligram per liter (mg/L) \\
		$V$ & Volume & $[L^3]$ & cubic feet ($\text{ft}^3$)  \\
		$v$ & Flow velocity & $[LT^{-1}]$ & feet per second (ft/sec)  \\
		\toprule \bottomrule
	\end{tabular}}
\end{table}

\section{Water Quality Controllability Dependency on System Hydraulics---An Example}\label{sec:App1}

In this appendix, we demonstrate the dependence of water quality controllability on system hydraulics using a straightforward example of the Three-node network. In our example, flow directions are assumed to be as illustrated in Fig. \ref{fig:CaseStudy}. Tank TK1 is considered filling and off-demand and a booster station is located at Junction J1 dosing chlorine into the system at a rate of $q_1^{\mathrm{B}}(t)$. The WQ system matrices of Eq. \eqref{equ:WQSS} are as expressed in \eqref{eq:ExampleMatrices} for this scenario. Note that to calculate the concentrations at Junction J1 at time-step $t + \Delta t_\mathrm{WQ}$ following Eq. \eqref{equ:mb-junc}, the flow rates need to be at the same time-step. Nevertheless, the water quality time-step operates at the seconds/minutes scale, whereas the hydraulic time-step is on an hourly scale. That is, within the same hydraulic time-step, $q(t+\Delta t_\mathrm{WQ})=q(t)$ for all links.

\input{TechniquesExample}

The formulation of the water quality system matrices clearly demonstrates a strong dependency on the system hydraulics, which extends to the water quality controllability matrix and Gramian as well. However, when these matrices are multiplied to calculate the water quality controllability matrix and Gramian in \eqref{eq:control_matrix} and \eqref{equ:control_gram}, the nonlinearity order increases exponentially.

\balance
\bibliographystyle{IEEEtran}
\bibliography{WQHydro}
\end{document}

%% file: preamble_new.tex
\usepackage{enumitem}
\usepackage{balance}
\usepackage{epsfig,color,amsmath,cite}
\usepackage{amsthm} 
\usepackage{amsmath}    
\usepackage[T1]{fontenc}
\usepackage[utf8]{inputenc}
\usepackage{bm}
\usepackage{epstopdf}
\usepackage{amssymb}
\usepackage{url}
\usepackage{enumitem} 
\usepackage{multirow}
\usepackage{hhline}
\usepackage{booktabs}
\usepackage{mathtools}
\usepackage{makecell}

\usepackage[linesnumbered,boxed,commentsnumbered,ruled,vlined,longend]{algorithm2e}
\SetKwProg{Init}{Initialization}{}{Proceed}
\usepackage{comment}

\makeatother
\DeclareMathAlphabet\mathbfcal{OMS}{cmsy}{b}{n}

\newtheorem{mydef}{Definition}

\newtheorem{myrem}{Remark}



\usepackage{stackengine}

\newcommand{\mat}[1]{\boldsymbol{#1}}
\renewcommand{\vec}[1]{\boldsymbol{\mathrm{#1}}}

\providecommand{\mA}{\ensuremath{\mat{A}}}
\providecommand{\mB}{\ensuremath{\mat{B}}}
\providecommand{\mC}{\ensuremath{\mat{C}}}

\providecommand{\mE}{\ensuremath{\mat{E}}}

\providecommand{\mT}{\ensuremath{\mat{T}}}

\providecommand{\vc}{\ensuremath{\vec{c}}}

\providecommand{\vl}{\ensuremath{\vec{l}}}

\providecommand{\vs}{\ensuremath{\vec{s}}}

\providecommand{\vu}{\ensuremath{\vec{u}}}
\providecommand{\vw}{\ensuremath{\vec{w}}}
\providecommand{\vx}{\ensuremath{\vec{x}}}
\providecommand{\vy}{\ensuremath{\vec{y}}}
\providecommand{\vz}{\ensuremath{\vec{z}}}

\newcommand{\m}{\boldsymbol}
\allowdisplaybreaks[4]
\usepackage[colorlinks = true,
linkcolor = blue,
urlcolor  = blue,
citecolor = blue,
anchorcolor = blue]{hyperref}

\newcommand{\mr}[1]{\mathrm{#1}}
\usepackage[framemethod=TikZ]{mdframed}
\mdfdefinestyle{MyFrame}{%
	linecolor=black,
	outerlinewidth=1.25pt,
	roundcorner=1.25pt,
	innerrightmargin=5pt,
	innerleftmargin=5pt,}
	
\usepackage{tikz}
\usetikzlibrary{matrix,positioning,decorations.pathreplacing}
\usetikzlibrary{shapes.geometric, arrows}
\usetikzlibrary{backgrounds}
\usetikzlibrary{shapes}
\usetikzlibrary{tikzmark}
\usetikzlibrary{calc}
\usetikzlibrary{arrows,shapes,positioning,shadows,trees,mindmap}
\usepackage[edges]{forest}
\usetikzlibrary{arrows.meta}
\colorlet{linecol}{black!75}
\usepackage{xkcdcolors} 

\usepackage[noabbrev]{cleveref}

\usepackage{mathtools}

\DeclarePairedDelimiter\abs{\lvert}{\rvert}%
\DeclarePairedDelimiter\norm{\lVert}{\rVert}%

\makeatletter
\let\oldabs\abs
\def\abs{\@ifstar{\oldabs}{\oldabs*}}
\let\oldnorm\norm
\def\norm{\@ifstar{\oldnorm}{\oldnorm*}}
\makeatother

\usepackage[english]{babel}
\usepackage[utf8]{inputenc}
\usepackage[super]{nth}

\usepackage{graphicx}
\usepackage{float}
\usepackage[caption = false]{subfig}

\usepackage{array}
\usepackage{threeparttable}

\usepackage[english]{babel}
\usepackage[utf8]{inputenc}
\usepackage[super]{nth}

\SetKwRepeat{Do}{do}{while}%

\everymath{\displaystyle}



%% file: TechniquesExample.tex
\begin{subequations}~\label{eq:ExampleMatrices}
	\begin{align}
\setlength{\tabcolsep}{1pt}
\mA_{\mathrm{WQ}}(t) & = 
{\small\begin{tabular}{c c}
			$\begin{bmatrix} 1 & 0 & 0 & 0 & 0 & 0 & 0 \\
			0 & 0 & 0 & \textcolor{violet}{{a}_{{\mathrm{J}}}} & 0 & 0 & 0 \\ 
			0 & 0 & \textcolor{violet}{{a}_{{\mathrm{TK}}}} & 0 & 0 & 0 & \textcolor{violet}{\underline{a}_{{\mathrm{TK}}}} \\ 
			1 & 0 & 0 & 0 & 0 & 0 & 0 \\
			0 & \textcolor{violet}{\underline{a}_{{\mathrm{P}}}} & 0 & 0 & \textcolor{violet}{{a}_{{\mathrm{P}}}} & 0 & 0 \\
			0 & 0 & 0 & 0 & \textcolor{violet}{\underline{a}_{{\mathrm{P}}}} & \textcolor{violet}{{a}_{{\mathrm{P}}}} & 0 \\
			0 & 0 & 0 & 0 & 0 & \textcolor{violet}{\underline{a}_{{\mathrm{P}}}} & \textcolor{violet}{{a}_{{\mathrm{P}}}}
		\end{bmatrix} $	 & $\begin{matrix}
		\begin{array}{c}
			$\textcolor{gray}{R1}$\\			
			$\textcolor{gray}{J1}$\\
			$\textcolor{gray}{TK1}$\\
			$\textcolor{gray}{M1}$\\
			\\
			$\textcolor{gray}{P1}$
			\\ \mathbb{ }
		\end{array}
	\end{matrix} , $
\end{tabular}}\\ 
\mB_{\mathrm{WQ}}(t) & =
{\small\begin{tabular}{c c}
		$\begin{bmatrix} 0 & 0  \\
			\textcolor{violet}{{a}^\mathrm{B}_{{\mathrm{J}}}} & 0 \\ 
			0 & 0  \\ 
			0 & 0  \\
			0 & 0  \\
			0 & 0  \\
			0 & 0 
		\end{bmatrix} $	 & $\begin{matrix}
			\begin{array}{c}
				$\textcolor{gray}{R1}$\\			
				$\textcolor{gray}{J1}$\\
				$\textcolor{gray}{TK1}$\\
				$\textcolor{gray}{M1}$\\
				\\
				$\textcolor{gray}{P1}$
				\\ \mathbb{ }
			\end{array}
		\end{matrix} ,$
\end{tabular}}
	\end{align}
\end{subequations}
where
\begin{subequations}
	\begin{align*}
	&	\textcolor{violet}{{a}_{{\mathrm{J}}}} = \frac{\textcolor{violet}{q_1^{\mathrm{M}}(t)}}{\textcolor{violet}{q^{\mathrm{D}_\mathrm{J}}_1(t)}+\textcolor{violet}{q^{\mathrm{P}}_{1}(t)}}, &
		\textcolor{violet}{{a}^\mathrm{B}_{{\mathrm{J}}}}=\frac{\textcolor{violet}{q_1^{\mathrm{B}}(t)}}{\textcolor{violet}{q^{\mathrm{D}_\mathrm{J}}_1(t)}+\textcolor{violet}{q^{\mathrm{P}}_{1}(t)}},  \\
		& \textcolor{violet}{{a}_{{\mathrm{TK}}}} = \frac{(1-k_b)\textcolor{violet}{V_1^\mathrm{TK}(t)} }{\textcolor{violet}{V_1^\mathrm{TK}(t+\Delta t_\mathrm{WQ})}},
		& \textcolor{violet}{\underline{a}_{{\mathrm{TK}}}} = \frac{\textcolor{violet}{q^{\mathrm{P}}_{1}(t) \Delta t_\mathrm{WQ}}}{\textcolor{violet}{V_1^\mathrm{TK}(t+\Delta t_\mathrm{WQ})}}, \\
		& \textcolor{violet}{{a}_{{\mathrm{P}}}} = 1 - \textcolor{violet}{\lambda(t)} - k_b \Delta t_\mathrm{WQ},
		& \textcolor{violet}{\underline{a}_{{\mathrm{P}}}}  = \textcolor{violet}{\lambda(t)}.
	\end{align*}
\end{subequations}